\documentclass[10pt,journal]{IEEEtran}

\usepackage{cite}
\usepackage[pdftex]{graphicx}
\DeclareGraphicsExtensions{.pdf,.png,.jpg,.jpeg} 
\graphicspath{{figures/}{pictures/}{images/}{./}} 
\usepackage{amsmath,amssymb,amsfonts}
\usepackage{algorithm,algpseudocode}
\usepackage{booktabs,multirow}
\usepackage{url,hyperref}
\makeatletter
\newcommand{\setword}[2]{%
  \phantomsection
  #1\def\@currentlabel{\unexpanded{#1}}\label{#2}%
} 
\makeatother
\usepackage{balance}
\usepackage{textcomp}
\usepackage{bm}

\usepackage{xcolor}
\definecolor{mycolor1}{RGB}{252,227,189}
\definecolor{mycolor2}{RGB}{120,191,157}
\definecolor{mycolor3}{RGB}{186,225,232}
\definecolor{mycolor4}{RGB}{193,152,131}
\definecolor{mycolor5}{RGB}{226,107,87}

\def\V{{\bm V}}
\def\hV{{\hat{\V}}}

\def\bsigma{{\bm \sigma}}
\def\eg{{\textit{e}.\textit{g}.}}
\def\ie{{\textit{i}.\textit{e}.}}
\def\etal{{\textit{et al}.}}

\usepackage{tikz}
\usetikzlibrary{backgrounds,calc,chains,fit,matrix,positioning, arrows.meta,shadows,shapes.misc,circuits.ee,shapes}
\tikzset{input/.style={}}
\tikzset{output/.style={}}
\tikzset{operator/.style={circle, draw, fill=black!8, minimum size=2.5ex, inner sep=0pt}}
\tikzset{filter/.style={rectangle, draw, fill=black!8, minimum size=3.5ex, inner xsep=1.5ex, align=center}}
\tikzset{filter1/.style={rectangle, draw, dashed, fill=black!8, minimum size=3.5ex, inner xsep=1.5ex, align=center}}
\tikzset{other/.style={rounded rectangle, draw, fill=black!8, minimum size=3.5ex, inner xsep=1ex}}
\tikzset{branch/.style={circle, draw, fill=black, minimum size=.5ex, inner sep=0pt}}
\tikzset{rv/.style={circle, draw, thick, fill=white, minimum size=2.75ex, inner sep=0pt}}
\tikzset{ob/.style={circle, draw, thick, fill=lightgray, minimum size=2.75ex, inner sep=0pt}}
\tikzset{pa/.style={circle, draw, thick, fill=black, minimum size=1ex, inner sep=0pt}}
\tikzset{/tikz/thin/.style={line width=.9pt}}
\tikzset{/tikz/thick/.style={line width=1.4pt}}
\tikzset{every path/.style={thin}}
\tikzset{>=direction ee}
\tikzset{every loop/.style={min distance=10mm,in=-60,out=60,looseness=10}}
\usepackage{pgfplots}
\pgfplotsset{compat=1.14}
\pgfplotsset{every axis/.append style={enlargelimits={abs=3pt},grid,axis lines=left}}
\pgfplotsset{every axis plot/.append style={thick,mark size=1.5pt,line join=bevel,mark options={solid}}}
\pgfplotsset{label style={font=\small}}
\pgfplotsset{tick label style={font=\footnotesize}}
\pgfplotsset{grid style={color=black!10}}
\pgfplotsset{legend style={draw=none,opacity=.85,font=\footnotesize,cells={anchor=west,opacity=1}}}
\pgfplotsset{every non boxed x axis/.style={xtick align=center,shorten <=-.5\pgflinewidth}}
\pgfplotsset{every non boxed y axis/.style={ytick align=center,shorten <=-.5\pgflinewidth}}
\pgfplotsset{every non boxed z axis/.style={ztick align=center,shorten <=-.5\pgflinewidth}}
\pgfplotsset{/pgf/number format/1000 sep={\,}}

\newcommand{\occupiedvoxels}[6]{
  \begin{scope}[xshift=2.25*#1cm,yshift=2.25*#2cm]
  \ifthenelse{#3 = 1}
  {\filldraw [fill=black!64, line width=1.6pt] (-0.75,-0.75) circle (0.5cm)}
  {\ifthenelse{#3 = 0}
  {\filldraw [fill=white, line width=1.6pt] (-0.75,-0.75) circle (0.5cm)}
  {\node at (-0.75,-0.75) (unknown) {\scriptsize $\bm{?}$}};
  }; 
  \ifthenelse{#4 = 1}
  {\filldraw [fill=black!64, line width=1.6pt] (-0.75,0.75) circle (0.5cm)}
  {\ifthenelse{#4 = 0}
  {\filldraw [fill=white, line width=1.6pt] (-0.75,0.75) circle (0.5cm)}
  {\node at (-0.75,0.75) (unknown) {\scriptsize $\bm{?}$}};
  }; 
  \ifthenelse{#5 = 1}
  {\filldraw [fill=black!64, line width=1.6pt] (0.75,-0.75) circle (0.5cm)}
  {\ifthenelse{#5 = 0}
  {\filldraw [fill=white, line width=1.6pt] (0.75,-0.75) circle (0.5cm)}
  {\node at (0.75,-0.75) (unknown) {\scriptsize $\bm{?}$}};
  }; 
  \ifthenelse{#6 = 1}
  {\filldraw [fill=black!64, line width=1.6pt] (0.75,0.75) circle (0.5cm)}
  {\ifthenelse{#6 = 0}
  {\filldraw [fill=white, line width=1.6pt] (0.75,0.75) circle (0.5cm)}
  {\node at (0.75,0.75) (unknown) {\scriptsize $\bm{?}$}};
  };
  \end{scope}
}
\newcommand{\occupieduniquevoxel}[2]{
  \begin{scope}[xshift=2.25*#1cm,yshift=2.25*#2cm]
  \filldraw [fill=black!64, line width=1.6pt] (0,0) circle (0.5cm); 
  \end{scope}
}
\newcommand{\occupiedtwoyvoxels}[4]{
  \begin{scope}[xshift=2.25*#1cm,yshift=2.25*#2cm]
  \ifthenelse{#3 = 1}
  {\filldraw [fill=black!64, line width=1.6pt] (0,-0.75) circle (0.5cm)}
  {\ifthenelse{#3 = 0}
  {\filldraw [fill=white, line width=1.6pt] (0,-0.75) circle (0.5cm)}
  {\node at (0,-0.75) (unknown) {\scriptsize $\bm{?}$}};
  }; 
  \ifthenelse{#4 = 1}
  {\filldraw [fill=black!64, line width=1.6pt] (0,0.75) circle (0.5cm)}
  {\ifthenelse{#4 = 0}
  {\filldraw [fill=white, line width=1.6pt] (0,0.75) circle (0.5cm)}
  {\node at (0,0.75) (unknown) {\scriptsize $\bm{?}$}};
  }; 
  \end{scope}
}
\newcommand{\occupiedtwoxvoxels}[4]{
  \begin{scope}[xshift=2.25*#1cm,yshift=2.25*#2cm]
  \ifthenelse{#3 = 1}
  {\filldraw [fill=black!64, line width=1.6pt] (-0.75,0) circle (0.5cm)}
  {\ifthenelse{#3 = 0}
  {\filldraw [fill=white, line width=1.6pt] (-0.75,0) circle (0.5cm)}
  {\node at (-0.75,0) (unknown) {\scriptsize $\bm{?}$}};
  }; 
  \ifthenelse{#4 = 1}
  {\filldraw [fill=black!64, line width=1.6pt] (0.75,0) circle (0.5cm)}
  {\ifthenelse{#4 = 0}
  {\filldraw [fill=white, line width=1.6pt] (0.75,0) circle (0.5cm)}
  {\node at (0.75,0) (unknown) {\scriptsize $\bm{?}$}};
  }; 
  \end{scope}
}
\newcommand{\occupiedtwoxvoxelswitherror}[4]{
  \begin{scope}[xshift=2.25*#1cm,yshift=2.25*#2cm]
  \ifthenelse{#3 = 1}
  {\filldraw [fill=black!64, line width=1.6pt] (-0.75,0) circle (0.5cm)}
  {\ifthenelse{#3 = 0}
  {\filldraw [fill=white, line width=1.6pt] (-0.75,0) circle (0.5cm)}
  {\filldraw [fill=red, line width=1.6pt] (-0.75,0) circle (0.5cm)};
  }; 
  \ifthenelse{#4 = 1}
  {\filldraw [fill=black!64, line width=1.6pt] (0.75,0) circle (0.5cm)}
  {\ifthenelse{#4 = 0}
  {\filldraw [fill=white, line width=1.6pt] (0.75,0) circle (0.5cm)}
  {\filldraw [fill=red, line width=1.6pt] (0.75,0) circle (0.5cm)};
  }; 
  \end{scope}
}
\newcommand{\occupiedvoxelswitherror}[6]{
  \begin{scope}[xshift=2.25*#1cm,yshift=2.25*#2cm]
  \ifthenelse{#3 = 1}
  {\filldraw [fill=black!64, line width=1.6pt] (-0.75,-0.75) circle (0.5cm)}
  {\ifthenelse{#3 = 0}
  {\filldraw [fill=white, line width=1.6pt] (-0.75,-0.75) circle (0.5cm)}
  {\filldraw [fill=red, line width=1.6pt] (-0.75,-0.75) circle (0.5cm)};
  }
  ; 
  \ifthenelse{#4 = 1}
  {\filldraw [fill=black!64, line width=1.6pt] (-0.75,0.75) circle (0.5cm)}
  {\ifthenelse{#4 = 0}
  {\filldraw [fill=white, line width=1.6pt] (-0.75,0.75) circle (0.5cm)}
  {\filldraw [fill=red, line width=1.6pt] (-0.75,0.75) circle (0.5cm)};
  }
  ; 
  \ifthenelse{#5 = 1}
  {\filldraw [fill=black!64, line width=1.6pt] (0.75,-0.75) circle (0.5cm)}
  {\ifthenelse{#5 = 0}
  {\filldraw [fill=white, line width=1.6pt] (0.75,-0.75) circle (0.5cm)}
  {\filldraw [fill=red, line width=1.6pt] (0.75,-0.75) circle (0.5cm)};
  }
  ; 
  \ifthenelse{#6 = 1}
  {\filldraw [fill=black!64, line width=1.6pt] (0.75,0.75) circle (0.5cm)}
  {\ifthenelse{#6 = 0}
  {\filldraw [fill=white, line width=1.6pt] (0.75,0.75) circle (0.5cm)}
  {\ifthenelse{#6 = -1}
  {\filldraw [fill=red, line width=1.6pt] (0.75,0.75) circle (0.5cm)}
  {\filldraw [fill=green, line width=1.6pt] (0.75,0.75) circle (0.5cm)};
  };
  };
  \end{scope}
}
\newcommand{\occupiedvoxelsmall}[3]{
  \begin{scope}[xshift=2.25*#1cm,yshift=2.25*#2cm]
  \node at (0,0) [circle] (v) {};
  \ifthenelse{#3 = 1}
  {\node [rectangle,draw,fit=(v),fill=black!32,inner sep=0,line width=1.6pt, minimum size=0.5cm] {}}
  {\node [rectangle,draw,fit=(v),fill=white,inner sep=0,line width=1.6pt, minimum size=0.5cm] {}};
  \end{scope}
}
\newcommand{\occupiedvoxel}[3]{
  \begin{scope}[xshift=2.25*#1cm,yshift=2.25*#2cm]
  \node at (0,0) [circle] (v) {};
  \ifthenelse{#3 = 1}
  {\node [rectangle,draw,fit=(v),fill=black!32,inner sep=0,line width=1.6pt, minimum size=1cm] {}}
  {\node [rectangle,draw,fit=(v),fill=white,inner sep=0,line width=1.6pt, minimum size=1cm] {}};
  \end{scope}
}

\newcommand{\anotherdemo}{
\begin{tikzpicture}[scale=0.25]
\begin{scope}[line cap=round]
\occupiedvoxel{0}{0}{0}
\occupiedvoxel{0}{2}{1}
\occupiedvoxels{0}{2}{0}{0}{1}{1}
\occupiedvoxel{0}{4}{1}
\occupiedvoxels{0}{4}{0}{0}{1}{1}
\occupiedvoxel{0}{6}{0}
\occupiedvoxel{2}{0}{1}
\occupiedvoxels{2}{0}{0}{0}{0}{1}
\occupiedvoxel{2}{2}{1}
\occupiedvoxels{2}{2}{1}{1}{1}{1}
\occupiedvoxel{2}{4}{1}
\occupiedvoxels{2}{4}{1}{1}{1}{1}
\occupiedvoxel{2}{6}{1}
\occupiedvoxels{2}{6}{1}{1}{1}{0}
\occupiedvoxel{4}{0}{1}
\occupiedvoxels{4}{0}{0}{1}{0}{1}
\occupiedvoxel{4}{2}{1}
\occupiedvoxels{4}{2}{1}{1}{1}{1}
\occupiedvoxel{4}{4}{1}
\occupiedvoxels{4}{4}{1}{1}{1}{0}
\occupiedvoxel{4}{6}{0}
\occupiedvoxel{6}{0}{0}
\occupiedvoxel{6}{2}{1}
\occupiedvoxels{6}{2}{1}{1}{1}{0}
\occupiedvoxel{6}{4}{0}
\occupiedvoxel{6}{6}{0}
\node at (16,6.5) (V) {};
\node at (29,6.5) (Vc) {};
\draw[->] (V)--(Vc) node[midway] (nd) {};
\node [above=1pt of nd] {\large 
$\hV^{(k)}\to\left\{\hV^{(k)}_{r}\right\}$};
\node [below=1pt of nd, align=center] {\large Based on \\ \large neighborhood \\ \large information};
\draw [decorate,decoration={brace,amplitude=10pt,raise=4pt},yshift=0pt]
(31,-2) -- (31,15);
\node at (32,13.5) (Vk0) {\large $\hV^{(k)}_{0}$: };
\node at (32,9) (Vk1) {\large $\hV^{(k)}_{1}$: };
\node at (32,4.5) (Vk2) {\large $\hV^{(k)}_{2}$: };
\node at (32,0) (Vk3) {\large $\hV^{(k)}_{3}$: };
\occupiedvoxel{16.2}{2}{1}
\occupiedvoxels{16.2}{2}{0}{0}{1}{1}
\occupiedvoxel{18.2}{2}{1}
\occupiedvoxels{18.2}{2}{0}{0}{1}{1}
\occupiedvoxel{20.2}{2}{1}
\occupiedvoxels{20.2}{2}{0}{0}{0}{1}
\occupiedvoxel{16.2}{0}{1}
\occupiedvoxels{16.2}{0}{1}{1}{1}{1}
\occupiedvoxel{18.2}{0}{1}
\occupiedvoxels{18.2}{0}{1}{1}{1}{1}
\occupiedvoxel{16.2}{6}{1}
\occupiedvoxels{16.2}{6}{1}{1}{1}{0}
\occupiedvoxel{18.2}{0}{1}
\occupiedvoxels{18.2}{0}{1}{1}{1}{1}
\occupiedvoxel{16.2}{4}{1}
\occupiedvoxels{16.2}{4}{0}{1}{0}{1}
\occupiedvoxel{20.2}{0}{1}
\occupiedvoxels{20.2}{0}{1}{1}{1}{1}
\occupiedvoxel{18.2}{4}{1}
\occupiedvoxels{18.2}{4}{1}{1}{1}{0}
\occupiedvoxel{20.2}{4}{1}
\occupiedvoxels{20.2}{4}{1}{1}{1}{0}
\node at (47,0) (Vk3r) {};
\node at (52,0) (sigmak3) {};
\draw[->] (Vk3r)--(sigmak3) node[midway] (ndk3) {};
\node [below=-4pt of ndk3, align=center] {\large Stats.};
\occupiedvoxel{24.2}{0}{1}
\occupiedvoxels{24.2}{0}{1}{1}{1}{1}
\node at (47,4.5) (Vk2r) {};
\node at (52,4.5) (sigmak2) {};
\draw[->] (Vk2r)--(sigmak2) node[midway] (ndk2) {};
\node [below=-4pt of ndk2, align=center] {\large Stats.};
\occupiedvoxel{24.2}{4}{1}
\occupiedvoxels{24.2}{4}{1}{1}{1}{0}
\node at (47,9) (Vk1r) {};
\node at (52,9) (sigmak1) {};
\draw[->] (Vk1r)--(sigmak1) node[midway] (ndk1) {};
\node [below=-4pt of ndk1, align=center] {\large Stats.};
\occupiedvoxel{24.2}{2}{1}
\occupiedvoxels{24.2}{2}{0}{0}{1}{1}
\node at (38,13.5) (Vk0r) {};
\node at (52,13.5) (sigmak0) {};
\draw[->] (Vk0r)--(sigmak0) node[midway] (ndk0) {};
\node [below=-4pt of ndk0, align=center] {\large Stats.};
\occupiedvoxel{24.2}{6}{1}
\occupiedvoxels{24.2}{6}{1}{1}{1}{0}
\draw [decorate,decoration={brace,amplitude=10pt,mirror,raise=4pt},yshift=0pt]
(56.2,-2) -- (56.2,15);
\node at (58,6.5) (Vkcr) {};
\node at (63,6.5) (sigmak) {};
\draw[->] (Vkcr)--(sigmak) node[midway] (ndsigmak) {};
\node [below=1pt of ndsigmak] {\large $\bsigma^{(k)}$};

\draw[->,ultra thick] (-3,-3)--(2,-3) node[right]{\large $x$};
\draw[->,ultra thick] (-3,-3)--(-3,2) node[above]{\large $y$};
\end{scope}
\end{tikzpicture}%
}

\newcommand{\demo}{
\begin{tikzpicture}[scale=0.15]
\begin{scope}[line cap=round]
\occupiedvoxelsmall{0}{0}{0}
\occupiedvoxelsmall{0}{2}{1}
\occupieduniquevoxel{0}{2}{1}
\occupiedvoxelsmall{0}{4}{1}
\occupiedtwoyvoxels{0}{4}{1}{1}
\occupiedvoxelsmall{0}{6}{1}
\occupieduniquevoxel{0}{6}{1}
\occupiedvoxelsmall{0}{8}{1}
\occupieduniquevoxel{0}{8}{1}
\occupiedvoxelsmall{0}{10}{0}
\occupiedvoxelsmall{2}{0}{0}
\occupiedvoxelsmall{2}{2}{1}
\occupieduniquevoxel{2}{2}{1}
\occupiedvoxelsmall{2}{4}{1}
\occupiedtwoyvoxels{2}{4}{1}{1}
\occupiedvoxelsmall{2}{6}{1}
\occupieduniquevoxel{2}{6}{1}
\occupiedvoxelsmall{2}{8}{1}
\occupieduniquevoxel{2}{8}{1}
\occupiedvoxelsmall{2}{10}{1}
\occupiedtwoyvoxels{2}{10}{1}{0}
\occupiedvoxelsmall{4}{0}{1}
\occupiedtwoxvoxels{4}{0}{1}{1}
\occupiedvoxelsmall{4}{2}{1}
\occupiedtwoxvoxels{4}{2}{1}{1}
\occupiedvoxelsmall{4}{4}{1}
\occupiedvoxels{4}{4}{1}{1}{1}{1}
\occupiedvoxelsmall{4}{6}{1}
\occupiedtwoxvoxels{4}{6}{1}{1}
\occupiedvoxelsmall{4}{8}{1}
\occupiedtwoxvoxels{4}{8}{1}{0}
\occupiedvoxelsmall{4}{10}{0}
\occupiedvoxelsmall{6}{0}{1}
\occupieduniquevoxel{6}{0}{1}
\occupiedvoxelsmall{6}{2}{1}
\occupieduniquevoxel{6}{2}{1}
\occupiedvoxelsmall{6}{4}{1}
\occupiedtwoyvoxels{6}{4}{1}{1}
\occupiedvoxelsmall{6}{6}{0}
\occupiedvoxelsmall{6}{8}{0}
\occupiedvoxelsmall{6}{10}{0}
\occupiedvoxelsmall{8}{0}{0}
\occupiedvoxelsmall{8}{2}{1}
\occupieduniquevoxel{8}{2}{1}
\occupiedvoxelsmall{8}{4}{1}
\occupiedtwoyvoxels{8}{4}{1}{0}
\occupiedvoxelsmall{8}{6}{0}
\occupiedvoxelsmall{8}{8}{0}
\occupiedvoxelsmall{8}{10}{0}
\node at (20,11) (V) {};
\node at (37,11) (Vc) {};
\draw[->] (V)--(Vc) node[midway] (nd) {};
\node [above=1pt of nd] {$\V^{(K)}\!\!\to\!\!\left\{\V^{(K)}_{c}\right\}$};
\node [below=1pt of nd, align=center] {Based on \\ coordinate \\ information};
\draw [decorate,decoration={brace,amplitude=10pt,raise=4pt},yshift=0pt]
(39,-2.5) -- (39,24.5);
\node at (41,22.5) (Vk0) {$\V^{(K)}_{0}$: };
\node at (41,15) (Vk1) {$\V^{(K)}_{1}$: };
\node at (41,6) (Vk2) {$\V^{(K)}_{2}$: };
\node at (41,-1) (Vk3) {$\V^{(K)}_{3}$: };
\occupiedvoxelsmall{20.7}{10}{1}
\occupieduniquevoxel{20.7}{10}{1}
\occupiedvoxelsmall{22.7}{10}{1}
\occupieduniquevoxel{22.7}{10}{1}
\occupiedvoxelsmall{24.7}{10}{1}
\occupieduniquevoxel{24.7}{10}{1}
\occupiedvoxelsmall{26.7}{10}{1}
\occupieduniquevoxel{26.7}{10}{1}
\occupiedvoxelsmall{28.7}{10}{1}
\occupieduniquevoxel{28.7}{10}{1}
\occupiedvoxelsmall{30.7}{10}{1}
\occupieduniquevoxel{30.7}{10}{1}
\occupiedvoxelsmall{32.7}{10}{1}
\occupieduniquevoxel{32.7}{10}{1}
\occupiedvoxelsmall{34.7}{10}{1}
\occupieduniquevoxel{34.7}{10}{1}
\occupiedvoxelsmall{36.7}{10}{1}
\occupieduniquevoxel{36.7}{10}{1}
\occupiedvoxelsmall{20.7}{6.75}{1}
\occupiedtwoxvoxels{20.7}{6.75}{1}{1}
\occupiedvoxelsmall{22.7}{6.75}{1}
\occupiedtwoxvoxels{22.7}{6.75}{1}{1}
\occupiedvoxelsmall{24.7}{6.75}{1}
\occupiedtwoxvoxels{24.7}{6.75}{1}{1}
\occupiedvoxelsmall{26.7}{6.75}{1}
\occupiedtwoxvoxels{26.7}{6.75}{1}{0}
\occupiedvoxelsmall{20.7}{2.75}{1}
\occupiedtwoyvoxels{20.7}{2.75}{1}{1}
\occupiedvoxelsmall{22.7}{2.75}{1}
\occupiedtwoyvoxels{22.7}{2.75}{1}{1}
\occupiedvoxelsmall{24.7}{2.75}{1}
\occupiedtwoyvoxels{24.7}{2.75}{1}{0}
\occupiedvoxelsmall{26.7}{2.75}{1}
\occupiedtwoyvoxels{26.7}{2.75}{1}{1}
\occupiedvoxelsmall{28.7}{2.75}{1}
\occupiedtwoyvoxels{28.7}{2.75}{1}{0}
\occupiedvoxelsmall{20.7}{-0.45}{1}
\occupiedvoxels{20.7}{-0.45}{1}{1}{1}{1}
\node at (47.5,-1) (VK3r) {};
\node at (85,-1) (sigmaK3) {};
\draw[->] (VK3r)--(sigmaK3) node[midway] (ndK3) {};
\node [above=-6pt of ndK3] {$\V^{(K)}_{3}\to\left\{\V^{(K)}_{3,r}\right\}$};
\node [below=-4pt of ndK3, align=center] {Based on neighborhood information};
\draw [decorate,decoration={brace,amplitude=10pt,raise=4pt},yshift=0pt]
(87,-2.75) -- (87,0.75);
\occupiedvoxelsmall{39.45}{-0.45}{1}
\occupiedvoxels{39.45}{-0.45}{1}{1}{1}{1}
\draw [decorate,decoration={brace,amplitude=10pt,mirror,raise=4pt},yshift=0pt]
(90,-2.75) -- (90,0.75);
\node at (93,-1) (VK3rr) {};
\node at (100,-1) (sigmaK3r) {};
\draw[->] (VK3rr)--(sigmaK3r) node[midway] (ndK3r) {};
\node [below=-4pt of ndK3r, align=center] {Stats.};
\draw [decorate,decoration={brace,amplitude=10pt,raise=4pt},yshift=0pt]
(103,-2.75) -- (103,0.75);
\occupiedvoxelsmall{46.45}{-0.45}{1}
\occupiedvoxels{46.45}{-0.45}{1}{1}{1}{1}
\draw [decorate,decoration={brace,amplitude=10pt,mirror,raise=4pt},yshift=0pt]
(105.5,-2.75) -- (105.5,0.75);
\node at (107.5,-1) (VK3rr) {};
\node at (115.5,-1) (sigmaK3r) {};
\draw[->] (VK3rr)--(sigmaK3r) node[midway] (ndK3r) {};
\node [below=-4pt of ndK3r] {$\bsigma^{(K)}_{3}$};
\node at (66.5,6) (VK2r) {};
\node at (75.5,6) (sigmaK2) {};
\draw[->] (VK2r)--(sigmaK2) node[midway] (ndK2) {};
\node [above=-6pt of ndK2] {$\left\{\V^{(K)}_{2,r}\right\}$};
\draw [decorate,decoration={brace,amplitude=10pt,raise=4pt},yshift=0pt]
(78.5,2.5) -- (78.5,9.5);
\node at (77.25,6.25) (VK2rr) {};
\node at (91.85,6.25) (sigmaK2r) {};
\draw[-] (VK2rr)--(sigmaK2r) node[midway] (ndK2r) {};
\node at (82.6,1.25) (VK2rc) {};
\node at (82.6,11.25) (sigmaK2c) {};
\draw[-] (VK2rc)--(sigmaK2c) node[midway] (ndK2c) {};
\draw [decorate,decoration={brace,amplitude=10pt,mirror,raise=4pt},yshift=0pt]
(90.5,2.5) -- (90.5,9.5);
\occupiedvoxelsmall{35.7}{3.75}{1}
\occupiedtwoyvoxels{35.7}{3.75}{1}{0}
\occupiedvoxelsmall{35.7}{1.75}{1}
\occupiedtwoyvoxels{35.7}{1.75}{1}{0}
\occupiedvoxelsmall{37.7}{3.75}{1}
\occupiedtwoyvoxels{37.7}{3.75}{1}{1}
\occupiedvoxelsmall{37.7}{1.75}{1}
\occupiedtwoyvoxels{37.7}{1.75}{1}{1}
\occupiedvoxelsmall{39.7}{1.75}{1}
\occupiedtwoyvoxels{39.7}{1.75}{1}{1}
\node at (93.5,6) (VK2rrr) {};
\node at (99,6) (sigmaK2rr) {};
\draw[->] (VK2rrr)--(sigmaK2rr) node[midway] (ndK2rr) {};
\node [below=-4pt of ndK2rr, align=center] {Stats.};
\draw [decorate,decoration={brace,amplitude=10pt,raise=4pt},yshift=0pt]
(102,2.5) -- (102,9.5);
\node at (100.75,6.25) (VK2rr) {};
\node at (110.75,6.25) (sigmaK2r) {};
\draw[-] (VK2rr)--(sigmaK2r) node[midway] (ndK2r) {};
\node at (106.1,1.25) (VK2rc) {};
\node at (106.1,11) (sigmaK2c) {};
\draw[-] (VK2rc)--(sigmaK2c) node[midway] (ndK2c) {};
\draw [decorate,decoration={brace,amplitude=10pt,mirror,raise=4pt},yshift=0pt]
(109.5,2.5) -- (109.5,9.5);
\occupiedvoxelsmall{46.15}{3.75}{1}
\occupiedtwoyvoxels{46.15}{3.75}{1}{0}
\occupiedvoxelsmall{46.15}{1.75}{1}
\occupiedtwoyvoxels{46.15}{1.75}{1}{0}
\occupiedvoxelsmall{48.15}{3.75}{1}
\occupiedtwoyvoxels{48.15}{3.75}{1}{1}
\occupiedvoxelsmall{48.15}{1.75}{1}
\occupiedtwoyvoxels{48.15}{1.75}{1}{1}
\node at (112,6) (VK2rr) {};
\node at (117.5,6) (sigmaK2r) {};
\draw[->] (VK2rr)--(sigmaK2r) node[midway] (ndK2r) {};
\node [below=-4pt of ndK2r] {$\bsigma^{(K)}_{2}$};
\node at (62,15) (Vk1r) {};
\node at (71,15) (sigmak1) {};
\draw[->] (Vk1r)--(sigmak1) node[midway] (ndK1) {};
\node [above=-6pt of ndK1] {$\left\{\V^{(K)}_{1,r}\right\}$};
\draw [decorate,decoration={brace,amplitude=10pt,raise=4pt},yshift=0pt]
(74,11.25) -- (74,18.75);
\draw [decorate,decoration={brace,amplitude=10pt,mirror,raise=4pt},yshift=0pt]
(85.5,11.25) -- (85.5,18.75);
\occupiedvoxelsmall{37.55}{7.65}{1}
\occupiedtwoxvoxels{37.55}{7.65}{1}{1}
\occupiedvoxelsmall{37.55}{5.6}{1}
\occupiedtwoxvoxels{37.55}{5.6}{1}{1}
\occupiedvoxelsmall{33.55}{5.6}{1}
\occupiedtwoxvoxels{33.55}{5.6}{1}{1}
\occupiedvoxelsmall{35.55}{5.6}{1}
\occupiedtwoxvoxels{35.55}{5.6}{1}{0}
\node at (72.5,14.9) (VK1rr) {};
\node at (87,14.9) (sigmaK1r) {};
\draw[-] (VK1rr)--(sigmaK1r) node[midway] (ndK1r) {};
\node at (82.2,10) (VK1rc) {};
\node at (82.2,19.75) (sigmaK1c) {};
\draw[-] (VK1rc)--(sigmaK1c) node[midway] (ndK1c) {};
\node at (88.5,15) (VK1rr) {};
\node at (94,15) (sigmaK1r) {};
\draw[->] (VK1rr)--(sigmaK1r) node[midway] (ndK1r) {};
\node [below=-4pt of ndK1r, align=center] {Stats.};
\draw [decorate,decoration={brace,amplitude=10pt,raise=4pt},yshift=0pt]
(97,11.25) -- (97,18.75);
\draw [decorate,decoration={brace,amplitude=10pt,mirror,raise=4pt},yshift=0pt]
(104.5,11.25) -- (104.5,18.75);
\occupiedvoxelsmall{45.95}{7.65}{1}
\occupiedtwoxvoxels{45.95}{7.65}{1}{1}
\occupiedvoxelsmall{45.95}{5.6}{1}
\occupiedtwoxvoxels{45.95}{5.6}{1}{1}
\occupiedvoxelsmall{43.95}{5.6}{1}
\occupiedtwoxvoxels{43.95}{5.6}{1}{1}
\node at (96,14.9) (VK1rr) {};
\node at (106,14.9) (sigmaK1r) {};
\draw[-] (VK1rr)--(sigmaK1r) node[midway] (ndK1r) {};
\node at (101.1,10) (VK1rc) {};
\node at (101.1,19.75) (sigmaK1c) {};
\draw[-] (VK1rc)--(sigmaK1c) node[midway] (ndK1c) {};
\node at (107,15) (Vk1rr) {};
\node at (113,15) (sigmak1r) {};
\draw[->] (Vk1rr)--(sigmak1r) node[midway] (ndk1r) {};
\node [below=-4pt of ndk1r] {$\bsigma^{(K)}_{1}$};
\draw [decorate,decoration={brace,amplitude=10pt,mirror,raise=4pt},yshift=0pt]
(115.25,-2) -- (115.25,19);
\node at (118,8.5) (Vkcr) {};
\node at (123,8.5) (sigmak) {};
\draw[->] (Vkcr)--(sigmak) node[midway] (ndsigmak) {};
\node [below=1pt of ndsigmak] {$\bsigma^{(K)}$};
\draw[->,ultra thick] (-3,-3)--(2,-3) node[right]{$x$};
\draw[->,ultra thick] (-3,-3)--(-3,2) node[above]{$y$};
\end{scope}
\end{tikzpicture}%
}

\newcommand{\idemo}{
\begin{tikzpicture}[scale=0.15]
\begin{scope}[line cap=round]
\node at (-0.5,30.4) (sigmaK) {$\bsigma^{(K)}$: };
\node at (7,30.4) (sigmaK1) {$\bsigma^{(K)}_1$};
\draw [decorate,decoration={brace,amplitude=10pt,raise=4pt},yshift=0pt]
(14.5,26.65) -- (14.5,34.15);
\draw [decorate,decoration={brace,amplitude=10pt,mirror,raise=4pt},yshift=0pt]
(21.5,26.65) -- (21.5,34.15);
\occupiedvoxelsmall{7}{12.5}{1}
\occupiedtwoxvoxels{7}{12.5}{1}{1}
\occupiedvoxelsmall{9}{12.5}{1}
\occupiedtwoxvoxels{9}{12.5}{1}{1}
\occupiedvoxelsmall{9}{14.5}{1}
\occupiedtwoxvoxels{9}{14.5}{1}{1}
\node at (12.75,30.4) (VK1rr) {};
\node at (23,30.4) (sigmaK1r) {};
\draw[-] (VK1rr)--(sigmaK1r) node[midway] (ndK1r) {};
\node at (18,25.4) (VK1rc) {};
\node at (18,35.4) (sigmaK1c) {};
\draw[-] (VK1rc)--(sigmaK1c) node[midway] (ndK1c) {};
\node at (25.5,30.4) (ddd) {,};
\node at (30,30.4) (sigmaK2) {$\bsigma^{(K)}_2$};
\occupiedvoxelsmall{17.2}{14.5}{1}
\occupiedtwoyvoxels{17.2}{14.5}{1}{0}
\occupiedvoxelsmall{17.2}{12.5}{1}
\occupiedtwoyvoxels{17.2}{12.5}{1}{0}
\occupiedvoxelsmall{19.2}{12.5}{1}
\occupiedtwoyvoxels{19.2}{12.5}{1}{1}
\occupiedvoxelsmall{19.2}{14.5}{1}
\occupiedtwoyvoxels{19.2}{14.5}{1}{1}
\draw [decorate,decoration={brace,amplitude=10pt,raise=4pt},yshift=0pt]
(37.5,26.65) -- (37.5,34.15);
\draw [decorate,decoration={brace,amplitude=10pt,mirror,raise=4pt},yshift=0pt]
(44.5,26.65) -- (44.5,34.15);
\node at (36,30.4) (VK2rr) {};
\node at (46.25,30.4) (sigmaK2r) {};
\draw[-] (VK2rr)--(sigmaK2r) node[midway] (ndK2r) {};
\node at (41,25.4) (VK2rc) {};
\node at (41,35.4) (sigmaK2c) {};
\draw[-] (VK2rc)--(sigmaK2c) node[midway] (ndK2c) {};
\node at (48.5,30.4) (ddd) {,};
\node at (53,30.4) (sigmaK3) {$\bsigma^{(K)}_3$};
\occupiedvoxelsmall{27.2}{13.5}{1}
\occupiedvoxels{27.2}{13.5}{1}{1}{1}{1}
\draw [decorate,decoration={brace,amplitude=10pt,raise=4pt},yshift=0pt]
(60,28.65) -- (60,32.4);
\draw [decorate,decoration={brace,amplitude=10pt,mirror,raise=4pt},yshift=0pt]
(62.5,28.65) -- (62.5,32.4);
\node at (78,30.4) (ddd) {used for interpolation.};
\occupiedvoxelsmall{0}{0}{0}
\occupiedvoxelsmall{0}{2}{1}
\occupieduniquevoxel{0}{2}{1}
\occupiedvoxelsmall{0}{4}{1}
\occupiedtwoyvoxels{0}{4}{-1}{-1}
\occupiedvoxelsmall{0}{6}{1}
\occupieduniquevoxel{0}{6}{1}
\occupiedvoxelsmall{0}{8}{1}
\occupieduniquevoxel{0}{8}{1}
\occupiedvoxelsmall{0}{10}{0}
\occupiedvoxelsmall{2}{0}{0}
\occupiedvoxelsmall{2}{2}{1}
\occupieduniquevoxel{2}{2}{1}
\occupiedvoxelsmall{2}{4}{1}
\occupiedtwoyvoxels{2}{4}{-1}{-1}
\occupiedvoxelsmall{2}{6}{1}
\occupieduniquevoxel{2}{6}{1}
\occupiedvoxelsmall{2}{8}{1}
\occupieduniquevoxel{2}{8}{1}
\occupiedvoxelsmall{2}{10}{1}
\occupiedtwoyvoxels{2}{10}{-1}{-1}
\occupiedvoxelsmall{4}{0}{1}
\occupiedtwoxvoxels{4}{0}{-1}{-1}
\occupiedvoxelsmall{4}{2}{1}
\occupiedtwoxvoxels{4}{2}{-1}{-1}
\occupiedvoxelsmall{4}{4}{1}
\occupiedvoxels{4}{4}{-1}{-1}{-1}{-1}
\occupiedvoxelsmall{4}{6}{1}
\occupiedtwoxvoxels{4}{6}{-1}{-1}
\occupiedvoxelsmall{4}{8}{1}
\occupiedtwoxvoxels{4}{8}{-1}{-1}
\occupiedvoxelsmall{4}{10}{0}
\occupiedvoxelsmall{6}{0}{1}
\occupieduniquevoxel{6}{0}{1}
\occupiedvoxelsmall{6}{2}{1}
\occupieduniquevoxel{6}{2}{1}
\occupiedvoxelsmall{6}{4}{1}
\occupiedtwoyvoxels{6}{4}{-1}{-1}
\occupiedvoxelsmall{6}{6}{0}
\occupiedvoxelsmall{6}{8}{0}
\occupiedvoxelsmall{6}{10}{0}
\occupiedvoxelsmall{8}{0}{0}
\occupiedvoxelsmall{8}{2}{1}
\occupieduniquevoxel{8}{2}{1}
\occupiedvoxelsmall{8}{4}{1}
\occupiedtwoyvoxels{8}{4}{-1}{-1}
\occupiedvoxelsmall{8}{6}{0}
\occupiedvoxelsmall{8}{8}{0}
\occupiedvoxelsmall{8}{10}{0}
\node at (20,11) (V) {};
\node at (35,11) (Vc) {};
\draw[->] (V)--(Vc) node[midway] (nd) {};
\node [above=1pt of nd] {\ $\V^{(K)}\!\!\to\!\!\left\{\V^{(K)}_{c}\right\}$};
\node [below=1pt of nd, align=center] {Based on \\ coordinate \\ information};
\draw [decorate,decoration={brace,amplitude=10pt,raise=4pt},yshift=0pt]
(38,-2.5) -- (38,24.5);
\node at (39,22.5) (Vk0) {$\V^{(K)}_{0}$: };
\node at (39,15) (Vk1) {$\V^{(K)}_{1}$};
\node at (39,6) (Vk2) {$\V^{(K)}_{2}$};
\node at (39,-1) (Vk3) {$\V^{(K)}_{3}$: };
\occupiedvoxelsmall{19.7}{10}{1}
\occupieduniquevoxel{19.7}{10}{1}
\occupiedvoxelsmall{21.7}{10}{1}
\occupieduniquevoxel{21.7}{10}{1}
\occupiedvoxelsmall{23.7}{10}{1}
\occupieduniquevoxel{23.7}{10}{1}
\occupiedvoxelsmall{25.7}{10}{1}
\occupieduniquevoxel{25.7}{10}{1}
\occupiedvoxelsmall{27.7}{10}{1}
\occupieduniquevoxel{27.7}{10}{1}
\occupiedvoxelsmall{29.7}{10}{1}
\occupieduniquevoxel{29.7}{10}{1}
\occupiedvoxelsmall{31.7}{10}{1}
\occupieduniquevoxel{31.7}{10}{1}
\occupiedvoxelsmall{33.7}{10}{1}
\occupieduniquevoxel{33.7}{10}{1}
\occupiedvoxelsmall{35.7}{10}{1}
\occupieduniquevoxel{35.7}{10}{1}
\occupiedvoxelsmall{20.7}{7.65}{1}
\occupiedtwoxvoxels{20.7}{7.65}{-1}{-1}
\occupiedvoxelsmall{22.7}{7.65}{1}
\occupiedtwoxvoxels{22.7}{7.65}{-1}{-1}
\occupiedvoxelsmall{20.7}{5.6}{1}
\occupiedtwoxvoxels{20.7}{5.6}{-1}{-1}
\occupiedvoxelsmall{22.7}{5.6}{1}
\occupiedtwoxvoxels{22.7}{5.6}{-1}{-1}
\occupiedvoxelsmall{20.7}{3.75}{1}
\occupiedtwoyvoxels{20.7}{3.75}{-1}{-1}
\occupiedvoxelsmall{22.7}{3.75}{1}
\occupiedtwoyvoxels{22.7}{3.75}{-1}{-1}
\occupiedvoxelsmall{24.7}{2.75}{1}
\occupiedtwoyvoxels{24.7}{2.75}{-1}{-1}
\occupiedvoxelsmall{20.7}{1.75}{1}
\occupiedtwoyvoxels{20.7}{1.75}{-1}{-1}
\occupiedvoxelsmall{22.7}{1.75}{1}
\occupiedtwoyvoxels{22.7}{1.75}{-1}{-1}
\occupiedvoxelsmall{19.7}{-0.45}{1}
\occupiedvoxels{19.7}{-0.45}{-1}{-1}{-1}{-1}
\draw [decorate,decoration={brace,amplitude=10pt,raise=4pt},yshift=0pt]
(45,11.25) -- (45,18.75);
\draw [decorate,decoration={brace,amplitude=10pt,raise=4pt},yshift=0pt]
(45,2.5) -- (45,9.5);
\node at (45.5,-1) (VK3r) {};
\node at (83,-1) (sigmaK3) {};
\draw[->] (VK3r)--(sigmaK3) node[midway] (ndK3) {};
\node [above=-6pt of ndK3] {Based on neighborhood information};
\node [below=-4pt of ndK3, align=center] {Divide $\V^{(K)}_{3}$ into $\left\{\V^{(K)}_{3,r}\right\}$};
\draw [decorate,decoration={brace,amplitude=10pt,raise=4pt},yshift=0pt]
(85,-2.75) -- (85,0.75);
\occupiedvoxelsmall{38.45}{-0.45}{1}
\occupiedvoxels{38.45}{-0.45}{-1}{-1}{-1}{-1}
\draw [decorate,decoration={brace,amplitude=10pt,mirror,raise=4pt},yshift=0pt]
(88,-2.75) -- (88,0.75);
\node at (91,-1) (VK3rr) {};
\node at (104,-1) (sigmaK3r) {};
\draw[->] (VK3rr)--(sigmaK3r) node[midway] (ndK3r) {};
\node [above=-6pt of ndK3r, align=center] {Interpolation};
\node [below=-4pt of ndK3r, align=center] {with $\bsigma^{(K)}_{3}$};
\draw [decorate,decoration={brace,amplitude=10pt,raise=4pt},yshift=0pt]
(107,-2.75) -- (107,0.75);
\occupiedvoxelsmall{48.15}{-0.45}{1}
\occupiedvoxels{48.15}{-0.45}{1}{1}{1}{1}
\node at (57,6) (VK2r) {};
\node at (66,6) (sigmaK2) {};
\draw[->] (VK2r)--(sigmaK2) node[midway] (ndK2) {};
\node [above=-6pt of ndK2] {$\left\{\V^{(K)}_{2,r}\right\}$};
\draw [decorate,decoration={brace,amplitude=10pt,raise=4pt},yshift=0pt]
(69,2.5) -- (69,9.5);
\node at (67.75,6.25) (VK2rr) {};
\node at (82.25,6.25) (sigmaK2r) {};
\draw[-] (VK2rr)--(sigmaK2r) node[midway] (ndK2r) {};
\node at (72.75,1.1) (VK2rc) {};
\node at (72.75,11.1) (sigmaK2c) {};
\draw[-] (VK2rc)--(sigmaK2c) node[midway] (ndK2c) {};
\draw [decorate,decoration={brace,amplitude=10pt,mirror,raise=4pt},yshift=0pt]
(81,2.5) -- (81,9.5);
\occupiedvoxelsmall{31.3}{3.75}{1}
\occupiedtwoyvoxels{31.3}{3.75}{-1}{-1}
\occupiedvoxelsmall{31.3}{1.75}{1}
\occupiedtwoyvoxels{31.3}{1.75}{-1}{-1}
\occupiedvoxelsmall{33.3}{3.75}{1}
\occupiedtwoyvoxels{33.3}{3.75}{-1}{-1}
\occupiedvoxelsmall{33.3}{1.75}{1}
\occupiedtwoyvoxels{33.3}{1.75}{-1}{-1}
\occupiedvoxelsmall{35.3}{1.75}{1}
\occupiedtwoyvoxels{35.3}{1.75}{-1}{-1}
\node at (84,6) (VK2rrr) {};
\node at (97,6) (sigmaK2rr) {};
\draw[->] (VK2rrr)--(sigmaK2rr) node[midway] (ndK2rr) {};
\node [above=-6pt of ndK2rr, align=center] {Interpolation};
\node [below=-4pt of ndK2rr, align=center] {with $\bsigma^{(K)}_{2}$};
\draw [decorate,decoration={brace,amplitude=10pt,raise=4pt},yshift=0pt]
(100,2.5) -- (100,9.5);
\node at (98.5,6.25) (VK2rr) {};
\node at (113,6.25) (sigmaK2r) {};
\draw[-] (VK2rr)--(sigmaK2r) node[midway] (ndK2r) {};
\node at (103.5,1.25) (VK2rc) {};
\node at (103.5,11) (sigmaK2c) {};
\draw[-] (VK2rc)--(sigmaK2c) node[midway] (ndK2c) {};
\occupiedvoxelsmall{45}{3.75}{1}
\occupiedtwoyvoxels{45}{3.75}{1}{0}
\occupiedvoxelsmall{45}{1.75}{1}
\occupiedtwoyvoxels{45}{1.75}{1}{0}
\occupiedvoxelsmall{47}{3.75}{1}
\occupiedtwoyvoxels{47}{3.75}{1}{1}
\occupiedvoxelsmall{47}{1.75}{1}
\occupiedtwoyvoxels{47}{1.75}{1}{1}
\occupiedvoxelsmall{49}{1.75}{1}
\occupiedtwoyvoxels{49}{1.75}{1}{1}
\node at (52,15) (Vk1r) {};
\node at (61.5,15) (sigmak1) {};
\draw[->] (Vk1r)--(sigmak1) node[midway] (ndK1) {};
\node [above=-6pt of ndK1] {$\left\{\V^{(K)}_{1,r}\right\}$};
\draw [decorate,decoration={brace,amplitude=10pt,raise=4pt},yshift=0pt]
(64,11.25) -- (64,18.75);
\draw [decorate,decoration={brace,amplitude=10pt,mirror,raise=4pt},yshift=0pt]
(75.5,11.25) -- (75.5,18.75);
\occupiedvoxelsmall{33}{7.65}{1}
\occupiedtwoxvoxels{33}{7.65}{-1}{-1}
\occupiedvoxelsmall{33}{5.6}{1}
\occupiedtwoxvoxels{33}{5.6}{-1}{-1}
\occupiedvoxelsmall{29}{5.6}{1}
\occupiedtwoxvoxels{29}{5.6}{-1}{-1}
\occupiedvoxelsmall{31}{5.6}{1}
\occupiedtwoxvoxels{31}{5.6}{-1}{-1}
\node at (62.5,14.9) (VK1rr) {};
\node at (77,14.9) (sigmaK1r) {};
\draw[-] (VK1rr)--(sigmaK1r) node[midway] (ndK1r) {};
\node at (72,10) (VK1rc) {};
\node at (72,19.75) (sigmaK1c) {};
\draw[-] (VK1rc)--(sigmaK1c) node[midway] (ndK1c) {};
\node at (78.5,15) (VK1rr) {};
\node at (91.5,15) (sigmaK1r) {};
\draw[->] (VK1rr)--(sigmaK1r) node[midway] (ndK1r) {};
\node [above=-6pt of ndK1r, align=center] {Interpolation};
\node [below=-4pt of ndK1r, align=center] {with $\bsigma^{(K)}_{1}$};
\draw [decorate,decoration={brace,amplitude=10pt,raise=4pt},yshift=0pt]
(94.5,11.25) -- (94.5,18.75);

\occupiedvoxelsmall{46.6}{7.65}{1}
\occupiedtwoxvoxels{46.6}{7.65}{1}{1}
\occupiedvoxelsmall{46.6}{5.6}{1}
\occupiedtwoxvoxels{46.6}{5.6}{1}{1}
\occupiedvoxelsmall{44.6}{5.6}{1}
\occupiedtwoxvoxels{44.6}{5.6}{1}{1}
\occupiedvoxelsmall{42.6}{5.6}{1}
\occupiedtwoxvoxels{42.6}{5.6}{1}{1}
\node at (93.25,14.9) (VK1rr) {};
\node at (107.75,14.9) (sigmaK1r) {};
\draw[-] (VK1rr)--(sigmaK1r) node[midway] (ndK1r) {};
\node at (102.6,10) (VK1rc) {};
\node at (102.6,19.75) (sigmaK1c) {};
\draw[-] (VK1rc)--(sigmaK1c) node[midway] (ndK1c) {};
\draw [decorate,decoration={brace,amplitude=10pt,mirror,raise=4pt},yshift=0pt]
(110.5,-4) -- (110.5,24);
\node at (113.5,10) (Vkcr) {};
\node at (121.5,10) (sigmak) {};
\draw[->] (Vkcr)--(sigmak) node[midway] (ndsigmak) {};
\node [below=1pt of ndsigmak] {$\hV^{(K-1)}$};
\draw[->,ultra thick] (-3,-3)--(2,-3) node[right]{$x$};
\draw[->,ultra thick] (-3,-3)--(-3,2) node[above]{$y$};
\occupiedvoxelsmall{55}{0}{0}
\occupiedvoxelsmall{55}{2}{1}
\occupieduniquevoxel{55}{2}{1}
\occupiedvoxelsmall{55}{4}{1}
\occupiedtwoyvoxels{55}{4}{1}{1}
\occupiedvoxelsmall{55}{6}{1}
\occupieduniquevoxel{55}{6}{1}
\occupiedvoxelsmall{55}{8}{1}
\occupieduniquevoxel{55}{8}{1}
\occupiedvoxelsmall{55}{10}{0}
\occupiedvoxelsmall{57}{0}{0}
\occupiedvoxelsmall{57}{2}{1}
\occupieduniquevoxel{57}{2}{1}
\occupiedvoxelsmall{57}{4}{1}
\occupiedtwoyvoxels{57}{4}{1}{1}
\occupiedvoxelsmall{57}{6}{1}
\occupieduniquevoxel{57}{6}{1}
\occupiedvoxelsmall{57}{8}{1}
\occupieduniquevoxel{57}{8}{1}
\occupiedvoxelsmall{57}{10}{1}
\occupiedtwoyvoxels{57}{10}{1}{0}
\occupiedvoxelsmall{59}{0}{1}
\occupiedtwoxvoxels{59}{0}{1}{1}
\occupiedvoxelsmall{59}{2}{1}
\occupiedtwoxvoxels{59}{2}{1}{1}
\occupiedvoxelsmall{59}{4}{1}
\occupiedvoxels{59}{4}{1}{1}{1}{1}
\occupiedvoxelsmall{59}{6}{1}
\occupiedtwoxvoxels{59}{6}{1}{1}
\occupiedvoxelsmall{59}{8}{1}
\occupiedtwoxvoxelswitherror{59}{8}{1}{2} 
\occupiedvoxelsmall{59}{10}{0}
\occupiedvoxelsmall{61}{0}{1}
\occupieduniquevoxel{61}{0}{1}
\occupiedvoxelsmall{61}{2}{1}
\occupieduniquevoxel{61}{2}{1}
\occupiedvoxelsmall{61}{4}{1}
\occupiedtwoyvoxels{61}{4}{1}{1}
\occupiedvoxelsmall{61}{6}{0}
\occupiedvoxelsmall{61}{8}{0}
\occupiedvoxelsmall{61}{10}{0}
\occupiedvoxelsmall{63}{0}{0}
\occupiedvoxelsmall{63}{2}{1}
\occupieduniquevoxel{63}{2}{1}
\occupiedvoxelsmall{63}{4}{1}
\occupiedtwoyvoxels{63}{4}{1}{0}
\occupiedvoxelsmall{63}{6}{0}
\occupiedvoxelsmall{63}{8}{0}
\occupiedvoxelsmall{63}{10}{0}
\end{scope}
\end{tikzpicture}%
}
\newcommand{\anotheridemo}{
\begin{tikzpicture}[scale=0.25]
\begin{scope}[line cap=round]
\occupiedvoxel{0}{0}{0}
\occupiedvoxel{0}{2}{1}
\occupiedvoxels{0}{2}{-1}{-1}{-1}{-1}
\occupiedvoxel{0}{4}{1}
\occupiedvoxels{0}{4}{-1}{-1}{-1}{-1}
\occupiedvoxel{0}{6}{0}
\occupiedvoxel{2}{0}{1}
\occupiedvoxels{2}{0}{-1}{-1}{-1}{-1}
\occupiedvoxel{2}{2}{1}
\occupiedvoxels{2}{2}{-1}{-1}{-1}{-1}
\occupiedvoxel{2}{4}{1}
\occupiedvoxels{2}{4}{-1}{-1}{-1}{-1}
\occupiedvoxel{2}{6}{1}
\occupiedvoxels{2}{6}{-1}{-1}{-1}{-1}
\occupiedvoxel{4}{0}{1}
\occupiedvoxels{4}{0}{-1}{-1}{-1}{-1}
\occupiedvoxel{4}{2}{1}
\occupiedvoxels{4}{2}{-1}{-1}{-1}{-1}
\occupiedvoxel{4}{4}{1}
\occupiedvoxels{4}{4}{-1}{-1}{-1}{-1}
\occupiedvoxel{4}{6}{0}
\occupiedvoxel{6}{0}{0}
\occupiedvoxel{6}{2}{1}
\occupiedvoxels{6}{2}{-1}{-1}{-1}{-1}
\occupiedvoxel{6}{4}{0}
\occupiedvoxel{6}{6}{0}
\node at (16,6.5) (V) {};
\node at (29,6.5) (Vc) {};
\draw[->] (V)--(Vc) node[midway] (nd) {};
\node [above=1pt of nd] {\large $\hV^{(k)}\to\left\{\hV^{(k)}_{r}\right\}$};
\node [below=1pt of nd, align=center] {\large Based on \\ \large neighborhood \\ \large information};
\draw [decorate,decoration={brace,amplitude=10pt,raise=4pt},yshift=0pt]
(31,-2) -- (31,15);
\node at (32,13.5) (Vk0) {\large $\hV^{(k)}_{0}$: };
\node at (32,9) (Vk1) {\large $\hV^{(k)}_{1}$: };
\node at (32,4.5) (Vk2) {\large $\hV^{(k)}_{2}$: };
\node at (32,0) (Vk3) {\large $\hV^{(k)}_{3}$: };
\occupiedvoxel{16.2}{2}{1}
\occupiedvoxels{16.2}{2}{-1}{-1}{-1}{-1}
\occupiedvoxel{18.2}{2}{1}
\occupiedvoxels{18.2}{2}{-1}{-1}{-1}{-1}
\occupiedvoxel{20.2}{2}{1}
\occupiedvoxels{20.2}{2}{-1}{-1}{-1}{-1}
\occupiedvoxel{16.2}{0}{1}
\occupiedvoxels{16.2}{0}{-1}{-1}{-1}{-1}
\occupiedvoxel{18.2}{0}{1}
\occupiedvoxels{18.2}{0}{-1}{-1}{-1}{-1}
\occupiedvoxel{16.2}{6}{1}
\occupiedvoxels{16.2}{6}{-1}{-1}{-1}{-1}
\occupiedvoxel{18.2}{0}{1}
\occupiedvoxels{18.2}{0}{-1}{-1}{-1}{-1}
\occupiedvoxel{16.2}{4}{1}
\occupiedvoxels{16.2}{4}{-1}{-1}{-1}{-1}
\occupiedvoxel{20.2}{0}{1}
\occupiedvoxels{20.2}{0}{-1}{-1}{-1}{-1}
\occupiedvoxel{18.2}{4}{1}
\occupiedvoxels{18.2}{4}{-1}{-1}{-1}{-1}
\occupiedvoxel{20.2}{4}{1}
\occupiedvoxels{20.2}{4}{-1}{-1}{-1}{-1}
\node at (47,0) (Vk3r) {};
\node at (61,0) (sigmak3) {};
\draw[->] (Vk3r)--(sigmak3) node[midway] (ndk3) {};
\node [below=-4pt of ndk3, align=center] {};
\occupiedvoxel{24.2}{1}{1}
\occupiedvoxels{24.2}{1}{1}{1}{1}{1}
\node at (47,4.5) (Vk2r) {};
\node at (61,4.5) (sigmak2) {};
\draw[->] (Vk2r)--(sigmak2) node[midway] (ndk2) {};
\node [below=-4pt of ndk2, align=center] {};
\occupiedvoxel{24.2}{5}{1}
\occupiedvoxels{24.2}{5}{1}{1}{1}{0}
\node at (47,9) (Vk1r) {};
\node at (61,9) (sigmak1) {};
\draw[->] (Vk1r)--(sigmak1) node[midway] (ndk1) {};
\node [below=-4pt of ndk1, align=center] {};
\occupiedvoxel{24.2}{3}{1}
\occupiedvoxels{24.2}{3}{0}{0}{1}{1}
\node at (38,13.5) (Vk0r) {};
\node at (61,13.5) (sigmak0) {};
\draw[->] (Vk0r)--(sigmak0) node[midway] (ndk0) {};
\node [above=4pt of ndk0] at ($(ndk0)+(-4,0)$) {\large Interpolation with};
\occupiedvoxel{24.2}{7}{1}
\occupiedvoxels{24.2}{7}{1}{1}{1}{0}
\node (sigmak) at (54.7,-2.5) {\large $\bsigma^{(k)}$};
\draw [decorate,decoration={brace,amplitude=10pt,mirror,raise=4pt},yshift=0pt]
(52.4,-0) -- (56.4,-0);
\draw [decorate,decoration={brace,amplitude=10pt,raise=4pt},yshift=0pt]
(52.4,18) -- (56.4,18);

\occupiedvoxel{28.2}{2}{1}
\occupiedvoxels{28.2}{2}{0}{0}{1}{1}
\occupiedvoxel{30.2}{2}{1}
\occupiedvoxels{30.2}{2}{0}{0}{1}{1}
\occupiedvoxel{32.2}{2}{1}
\occupiedvoxelswitherror{32.2}{2}{0}{0}{-1}{1} %
\occupiedvoxel{28.2}{0}{1}
\occupiedvoxels{28.2}{0}{1}{1}{1}{1}
\occupiedvoxel{30.2}{0}{1}
\occupiedvoxels{30.2}{0}{1}{1}{1}{1}
\occupiedvoxel{28.2}{6}{1}
\occupiedvoxels{28.2}{6}{1}{1}{1}{0}
\occupiedvoxel{30.2}{0}{1}
\occupiedvoxels{30.2}{0}{1}{1}{1}{1}
\occupiedvoxel{28.2}{4}{1}
\occupiedvoxelswitherror{28.2}{4}{-1}{1}{-1}{-2} %
\occupiedvoxel{32.2}{0}{1}
\occupiedvoxels{32.2}{0}{1}{1}{1}{1}
\occupiedvoxel{30.2}{4}{1}
\occupiedvoxels{30.2}{4}{1}{1}{1}{0}
\occupiedvoxel{32.2}{4}{1}
\occupiedvoxels{32.2}{4}{1}{1}{1}{0}

\draw [decorate,decoration={brace,amplitude=10pt,mirror,raise=4pt},yshift=0pt]
(74,-2) -- (74,15);
\node at (76,6.5) (Vkcr) {};
\node at (83,6.5) (sigmak) {};
\draw[->] (Vkcr)--(sigmak) node[midway] (ndsigmak) {};
\node [below=1pt of ndsigmak] {\large $\hV^{(k-1)}$};
\occupiedvoxel{38.2}{0}{0}
\occupiedvoxel{38.2}{2}{1}
\occupiedvoxels{38.2}{2}{0}{0}{1}{1}
\occupiedvoxel{38.2}{4}{1}
\occupiedvoxels{38.2}{4}{0}{0}{1}{1}
\occupiedvoxel{38.2}{6}{0}
\occupiedvoxel{40.2}{0}{1}
\occupiedvoxelswitherror{40.2}{0}{0}{0}{-1}{1}
\occupiedvoxel{40.2}{2}{1}
\occupiedvoxels{40.2}{2}{1}{1}{1}{1}
\occupiedvoxel{40.2}{4}{1}
\occupiedvoxels{40.2}{4}{1}{1}{1}{1}
\occupiedvoxel{40.2}{6}{1}
\occupiedvoxels{40.2}{6}{1}{1}{1}{0}
\occupiedvoxel{42.2}{0}{1}
\occupiedvoxelswitherror{42.2}{0}{-1}{1}{-1}{-2}
\occupiedvoxel{42.2}{2}{1}
\occupiedvoxels{42.2}{2}{1}{1}{1}{1}
\occupiedvoxel{42.2}{4}{1}
\occupiedvoxels{42.2}{4}{1}{1}{1}{0}
\occupiedvoxel{42.2}{6}{0}
\occupiedvoxel{44.2}{0}{0}
\occupiedvoxel{44.2}{2}{1}
\occupiedvoxels{44.2}{2}{1}{1}{1}{0}
\occupiedvoxel{44.2}{4}{0}
\occupiedvoxel{44.2}{6}{0}

\draw[->,ultra thick] (-3,-3)--(2,-3) node[right]{\large $x$};
\draw[->,ultra thick] (-3,-3)--(-3,2) node[above]{\large $y$};
\end{scope}
\end{tikzpicture}%
}
\newcommand{\hpsrpccDecoder}{
\begin{tikzpicture}[
    myfont/.style={
    },
    cell/.style={
        circle,draw, fill=black!8,
        line width = .75pt,
        minimum width=1cm,
        inner sep=1pt,
    },
    corner/.style={
        circle,draw, fill=black!8,
        line width = .75pt,
        inner sep=1pt,
    },
    bitstream/.style={
        ellipse,
        draw, very thick, fill=black!8,
        minimum width=1cm,
        minimum height=0.66cm,
        inner sep=1pt,
    },
    hpsr/.style={
        rectangle, draw, fill=black!8,
        rounded corners=5mm,
        minimum width=4cm,
        minimum height=2cm,
        inner sep=1pt,
        align=center, 
        very thick,
    },
    coder/.style={
        rectangle, draw, fill=black!8,
        minimum width=1.4cm,
        minimum height=1cm,
        inner sep=1pt,
        align=center, %
        very thick,
    },
    ArrowC1/.style={
        rounded corners=2mm,
        >=Stealth[round],
        very thick,
    },
]
    \node [hpsr] (hpsr) at (2.5,0) {\textbf{Hierarchical prior-based}\\ \textbf{super resolution}};
    \node[cell, label={[myfont]above:Reconstructed point cloud}] (rpc) at (-1.6,0) {$\hV$};
    \node[cell, label={[myfont]above:Base point cloud}] (bpc) at (6,0) {${\V}^{(K)}$};
    \node[cell, label={[myfont]below:Hierarchical prior}] (hp) at (6,-2) {$\left\{{\bsigma}^{(k)}\right\}_{k=1}^{K}$};
    \node [coder] (baseCoder) at (8.5,0) {Base \\ decoder};
    \node [coder] (sideCoder) at (8.5,-2) {Prior \\ decoder};
    \node[bitstream, label={[myfont]above:Base bitstream}] (basebit) at (10.5,0) {};
    \node[bitstream, label={[myfont]below:Prior bitstream}] (sidebit) at (10.5,-2) {};
    \draw [->, ArrowC1] (basebit) -- (baseCoder) -- (bpc) -- (hpsr);
    \draw [ArrowC1] (sidebit) -- (sideCoder) -- (hp);
    \draw [->, ArrowC1] (hp) -| (hpsr);
    \draw [->, ArrowC1] (hpsr) -- (rpc);   
\end{tikzpicture}
}
\newcommand{\hpsrpccEncoder}{
\begin{tikzpicture}[
    myfont/.style={
    },
    frame/.style={
        rectangle,
        rounded corners=5mm,
        draw,gray, fill=black!8,
        very thick,
    },
    cell/.style={
        circle,draw, fill=black!8,
        line width = .75pt,
        minimum width=1cm,
        inner sep=1pt,
    },
    corner/.style={
        circle,draw, fill=black!8,
        line width = .75pt,
        inner sep=1pt,
    },
    bitstream/.style={
        ellipse,
        draw,very thick, fill=black!8,
        minimum width=1cm,
        minimum height=0.66cm,
        inner sep=1pt,
    },
    downfunc/.style={
        rectangle, draw, fill=black!8,
        minimum width=1cm,
        minimum height=1cm,
        inner sep=1pt,
        very thick,
    },
    coder/.style={
        rectangle, draw, fill=black!8,
        minimum width=1.4cm,
        minimum height=1cm,
        inner sep=1pt,
        align=center, 
        very thick,
    },
    hpc/.style={
        rectangle, fill=black!8,
        rounded corners=5mm,
        draw,
        minimum width=4cm,
        minimum height=2cm,
        inner sep=1pt,
        align=center,
        very thick,
    },
    ArrowC1/.style={
        rounded corners=3mm,
        >=Stealth[round],
        very thick,
    },
]
    \node [frame, minimum height =3cm, minimum width=6cm, label={above:\textbf{Successive downsampling}}] at (0,0) {};
    \node [hpc] (hpc) at (1.8,-3) {\textbf{Hierarchical prior} \\ \textbf{construction}};
    
    \node [downfunc] (down0) at (-2,0) {$\downarrow s_0$};
    \node [downfunc] (down1) at (-0.5,0) {$\downarrow s_1$};
    \node [] (cdots) at (1,0) {$\cdots$};
    \node [downfunc] (downK) at (2,0) {$\downarrow s_K$};

    \node[cell, label={[myfont]above:Original point cloud}] (pc) at (-4.6,0) {{$\V$}};

    \node[cell, label={[myfont]above:Base point cloud}] (bpc) at (5,0) {${\V}^{(K)}$};
    \node[corner] (corner) at (-2,-1) {};
    \node[cell, label={[myfont]below:Point cloud pyramid}] (pcp) at (-2,-3) {$\left\{{\V}^{(k)}\right\}_{k=0}^{K}$};
    \node[cell, label={[myfont]below:Hierarchical prior}] (hp) at (5.5,-3) {$\left\{{\bsigma}^{(k)}\right\}_{k=1}^{K}$};

    \node [coder] (baseCoder) at (8,0) {Base \\ encoder};
    \node [coder] (sideCoder) at (8,-3) {Prior \\ encoder};
    
    \node[bitstream, label={[myfont]above:Base bitstream}] (basebit) at (10,0) {};
    \node[bitstream, label={[myfont]below:Prior bitstream}] (sidebit) at (10,-3) {};

    \draw [->, ArrowC1] (pc) -- (down0) -- (down1) -- (cdots) -- (downK) -- (bpc); 
    \draw [->, ArrowC1] (bpc) -- (baseCoder) -- (basebit);
    \draw [->, ArrowC1] (down0) -- (pcp);
    \draw [ArrowC1] (down1) |- (corner);
    \draw [dashed, ArrowC1] (cdots) |- (corner);
    \draw [ArrowC1] (downK) |- (corner);
    \draw [->, ArrowC1] (pcp) -- (hpc) -- (hp) -- (sideCoder) -- (sidebit);

\end{tikzpicture}
}
\newcommand{\hpcPipeline}{
\begin{tikzpicture}[x=2.8cm,y=1.5cm]
\node [input] (VK) {};
\node [above=6pt of VK] {$\V^{(K)}$};
\node [branch] (VKb) at ($(VK)+(0.15,0)$) {};
\node [filter] (hpcK) at ($(VK)+(1.1, 1)$) {Construction of $\bsigma^{(K)}$};
\node [input] (VKm1) at ($(hpcK)+(0, 0.75)$)  {};
\node [right=-4pt of VKm1] {$\V^{(K-1)}$};
\node [filter] (iK) at ($(hpcK)+(0,-1)$) {Super resolution of $\V^{(K)}$};
\draw[->] (VK) -- (iK);
\draw[->] (VKm1) -- (hpcK);
\draw[->,rounded corners=8pt] (VKb) |- ($(hpcK)+(-0.65,0)$);
\draw[->] (hpcK) -- (iK) node[midway] (sigmaK) {};
\node [right=-4pt of sigmaK] {$\bsigma^{(K)}$};
\node [branch] (VKm1b) at ($(iK)+(0.9,0)$) {};
\node [filter] (hpcKm1) at ($(hpcK)+(1.9, 0)$) {Construction of $\bsigma^{(K-1)}$};
\node [input] (VKm2) at ($(hpcKm1)+(0, 0.75)$)  {};
\node [right=-4pt of VKm2] {$\V^{(K-2)}$};
\node [filter] (iKm1) at ($(hpcKm1)+(0, -1)$) {Super resolution of $\hV^{(K-1)}$};
\draw[->] (iK) -- (iKm1) node[midway] (hVKm1) {};
\draw[->] (VKm2) -- (hpcKm1);
\draw[->,rounded corners=8pt] (VKm1b) |- ($(hpcKm1)+(-0.71,0)$);
\node [right=-4pt of hVKm1] at ($(hVKm1)+(0,0.5)$) {$\hV^{(K-1)}$};
\draw[->] (hpcKm1) -- (iKm1) node[midway] (sigmaKm1) {};
\node [right=-4pt of sigmaKm1] {$\bsigma^{(K-1)}$};
\node [input] (cdots) at ($(sigmaKm1)+(1.2,0)$) {$\cdots$};
\node [input] (V1i) at ($(iKm1)+(1.2,0)$) {};
\node [branch] (V1b) at ($(iKm1)+(1.5,0)$) {};
\node [filter] (hpc1) at ($(hpcKm1)+(2.5, 0)$) {Construction of $\bsigma^{(1)}$};
\node [input] (V0) at ($(hpc1)+(0, 0.75)$)  {};
\node [right=-4pt of V0] {$\V^{(0)}$};
\node [filter1] (i1) at ($(hpc1)+(0, -1)$) {Super resolution of $\hV^{(1)}$};
\draw (V1i) -- (V1b);
\draw[dashed,->] (V1i) -- (i1) node[midway] (hV1) {};
\draw[->] (V0) -- (hpc1);
\draw[->,rounded corners=8pt] (V1b) |- ($(hpc1)+(-0.63,0)$);
\node [right=0pt of V1b] at ($(V1b)+(0,0.5)$) {$\hV^{(1)}$};
\draw[dashed, ->] (hpc1) -- (i1) node[midway] (sigma1) {};
\node [right=-4pt of sigma1] {$\bsigma^{(1)}$};
\draw[dashed,->] ($(i1)+(0.75,0)$) -- ($(i1)+(1,0)$) node[midway] (hV0) {};
\node [above=6pt of hV0] {$\hV^{(0)}$};
\end{tikzpicture}%
}
\newcommand{\hpsr}{

\begin{tikzpicture}[x=2.8cm,y=1.5cm]
\node [input] (VK) {};
\node [above=-8pt of VK] {Base point cloud $\V^{(K)}$};
\node [filter] (iK) at ($(VK)+(0,-0.7)$) {Super resolution of $\V^{(K)}$\\ {with factor $1/\left(q\times 2^L\right)$}};
\draw[->] (VK) -- (iK);
\node [input] (hpcK) at ($(iK)+(-1.2, 0)$) {};
\draw[->] (hpcK) -- (iK) node[midway] (sigmaK) {};
\node [left=10pt of sigmaK] {$\bsigma^{(K)}$};
\node  [filter] (iKm) at ($(iK)+(0,-1.1)$) {Super resolution of $\hV^{(k)}$ \\ {with factor $2$} for $K>k\ge 1$};
\draw[->] (iK) -- (iKm) node[midway] (VKm1) {};
\node [right=-4pt of VKm1] {$\hV^{(K-1)}$};
\node [input] (hpcKm) at ($(iKm)+(-1.2, 0)$) {};
\draw[->] (hpcKm) -- (iKm) node[midway] (sigmaKm) {};
\node [left=10pt of sigmaKm] {$\bsigma^{(k)}$};
\node [branch] (hpcKm1) at ($(iKm)+(0.83, 0)$) {};
\path[-stealth] (hpcKm1) edge[loop=right] node[right] {$\hV^{(k)}$} ();
\node  [filter] (ir) at ($(iKm)+(0,-1.1)$) {{Super resolution} of $\hV^{(0)}$ \\ {with factor $2$} for $K'$ iterations};
\draw[->] (iKm) -- (ir) node[midway] (V0) {};
\node [right=-4pt of V0] {$\hV^{(0)}$};
\node [input] (hpc) at ($(ir)+(-1.2, 0)$) {};
\draw[->] (hpc) -- (ir) node[midway] (sigma) {};
\node [left=10pt of sigma] {$\bsigma^{(1)}$};
\node [branch] (hV0) at ($(ir)+(0.86, 0)$) {};
\path[-stealth] (hV0) edge[loop=right] node[right] {$\hV^{(0)}$} ();
\node  [filter] (iq) at ($(ir)+(0,-1.1)$) {Upscaling of $\hV^{(0)}$\\ {with factor $2^{L+1-K-K'}$}};
\draw[->] (ir) -- (iq) node[midway] (V0) {};
\node [right=-4pt of V0] {$\hV^{(0)}$};
\draw[->] (iq) -- ($(iq)+(0,-0.7)$) node[midway] (V) {};
\node [below=4pt of V] {{Reconstructed point cloud} $\hV$};
\end{tikzpicture}%
}
\hypersetup{hidelinks} %

\begin{document}
\title{Hierarchical Prior-based Super Resolution\\ for Point Cloud Geometry Compression}
\author{Dingquan Li, 
        Kede Ma, 
        Jing Wang, and 
        Ge Li
\thanks{D. Li and J. Wang are with the Network Intelligence Research Department, Peng Cheng Laboratory, Shenzhen, China (e-mail: dingquanli@pku.edu.cn; wangj@pcl.ac.cn).}
\thanks{K. Ma is with the Department of Computer Science, City University of Hong Kong, Hong Kong, China (e-mail: kede.ma@cityu.edu.hk).}
\thanks{G. Li is with the School of Electronic and Computer Engineering, Peking University Shenzhen Graduate School, Shenzhen, China (e-mail: geli@ece.pku.edu.cn).}%
}
\markboth{Preprint}
{Li~\MakeLowercase{\etal}: Hierarchical Prior-based Super Resolution for Point Cloud Geometry Compression}
\maketitle

\begin{abstract} 
The Geometry-based Point Cloud Compression (G-PCC) has been developed by the Moving Picture Experts Group to compress point clouds. In its lossy mode, the reconstructed point cloud by G-PCC often suffers from noticeable distortions due to the na\"{i}ve geometry quantization (\ie, grid downsampling). This paper proposes a hierarchical prior-based super resolution method for point cloud geometry compression. The content-dependent hierarchical prior is constructed at the encoder side, which enables coarse-to-fine super resolution of the point cloud geometry at the decoder side. A more accurate prior generally yields improved reconstruction performance, at the cost of increased bits required to encode this side information. With a proper balance between prior accuracy and bit consumption, the proposed method demonstrates substantial Bj{\o}ntegaard-delta bitrate savings on the MPEG Cat1A dataset, surpassing the octree-based and trisoup-based G-PCC v14. We provide our implementations for reproducible research at \url{https://github.com/lidq92/mpeg-pcc-tmc13}.
\end{abstract}

\begin{IEEEkeywords}
Point cloud geometry compression, hierarchical prior, coarse-to-fine super resolution
\end{IEEEkeywords}

\section{Introduction}\label{sec:introduction}
\IEEEPARstart{R}{ecent} advances in capturing and rendering 3D real-world scenes have expanded the frontiers of multimedia applications, offering immersive and interactive experiences. These applications, including virtual, augmented, and mixed reality, have become feasible through the remarkable progress in processing three-dimensional data~\cite{schwarz2019emerging}. Point clouds, among various means of representing 3D scenes and objects, emerge as a fundamental and primitive form. A point cloud comprises an unordered collection of points, each defined by spatial coordinates and accompanied by additional attributes such as color, reflectance, and surface normal. Point clouds possess distinctive advantages over alternative 3D data representations, such as polygonal meshes and multi-view images. Their inherent simplicity and flexibility enable efficient representation of non-manifold geometry without necessitating explicit connectivity information. Moreover, point clouds hold great potential for real-time rendering of high-quality visuals~\cite{akenine2018real}.

In various applications involving point clouds, such as cultural heritage preservation, 3D telepresence, and robotic navigation, it is often necessary to work with millions to billions of 3D points to achieve a high-quality representation with precise geometric details, typically at sub-centimeter precision. However, this poses a substantial challenge concerning storage, transmission, and manipulation. Point Cloud Compression (PCC) offers a solution by enabling users to interact with high-quality 3D point cloud content while alleviating the demands on storage and transmission compared to utilizing uncompressed raw data. Acknowledging the significance of PCC, the Moving Picture Experts Group (MPEG) has devoted considerable efforts to establish an open PCC standard~\cite{schwarz2019emerging}. In 2017, MPEG initiated a Call for Proposals on PCC, leading to the development of the first generation of MPEG PCC standard, comprising two classes of solutions: Video-based PCC (V-PCC) and Geometry-based PCC (G-PCC)~\cite{graziosi2020overview}. V-PCC utilizes 3D-to-2D projections to leverage existing video coding techniques for compression. In contrast, G-PCC directly operates on 3D point clouds by employing efficient data structures such as octrees~\cite{meagher1982geometric}. 

In data compression, lossy compression offers a valuable advantage over lossless compression by enabling a trade-off between the compression rate and distortion. This flexibility is particularly well-suited for scenarios with limited memory and bandwidth resources. The octree-based G-PCC approach implements lossy compression through na\"{i}ve geometry grid downsampling. This downsampling step, called geometry quantization in the MPEG G-PCC standard, results in noticeable distortions in the reconstructed point cloud. To overcome this limitation, Borges~\etal~\cite{borges2022fractional} introduced a post-hoc fractional super resolution technique called SRLUT, assuming cross-scale self-similarity. While post-processing techniques effectively reduce distortions without incurring additional bitrate costs, they generally fall short of optimizing rate-distortion performance.

In this paper, we introduce a Hierarchical Prior-based Super Resolution method for Point Cloud Geometry Compression (HPSR-PCGC). At the encoder side, we first create a pyramid of point clouds by successively downsampling the original point cloud $\V\in \mathbb{R}^{N\times 3}$, denoted as $\{\V^{(k)}\}_{k=0}^K$. With the assumption of non-local geometric similarity, we iteratively construct the hierarchical prior $\{{\bsigma}^{(k)}\}_{k=1}^{K}$ from the point cloud pyramid by neighborhood-based point clustering and frequency-based occupancy estimation. The final step of the encoder involves losslessly compressing both the base point cloud $\V^{(K)}$ and the hierarchical prior $\{{\bsigma}^{(k)}\}_{k=1}^{K}$ into separate bitstreams. At the decoder side, our method begins by decoding the bitstreams to reconstruct the base point cloud $\V^{(K)}$ and the hierarchical prior $\{{\bsigma}^{(k)}\}_{k=1}^{K}$. We then progressively interpolate the base point cloud $\V^{(K)}$ using the hierarchical prior $\{{\bsigma}^{(k)}\}_{k=1}^{K}$, resulting in the final reconstructed point cloud $\hV$.

We conduct experiments on the MPEG Cat1A dataset~\cite{mpeg2021ctc}, employing the octree-based G-PCC as the base encoder/decoder. The results demonstrate the effectiveness of our method, showcasing significant point-to-point (D1) and point-to-plane (D2) Bj{\o}ntegaard-delta bitrate savings, compared to the octree-based G-PCC, trisoup-based G-PCC, and SRLUT. 

\section{Related Work}\label{sec:related}
Our work centers on the intersection of point cloud geometry compression and super resolution, of which we provide a concise overview. 

\subsection{Point Cloud Geometry Compression}\label{subsec:pcgc}
\noindent\textbf{Traditional PCC}. 
Representative methods encompass V-PCC and G-PCC~\cite{cao2021compression}. For a static point cloud, V-PCC first segments it into a set of 3D patches, which are mapped onto a predefined set of 2D planes through orthogonal projections.  Patch packing is then executed on a regular 2D grid to create a 2D image representing the point cloud's geometry. A 2D occupancy map is also generated to identify grid cells containing the projected points. For a dynamic point cloud, a 2D geometry video and an occupancy video are generated and compressed using established video codecs, such as HEVC~\cite{sullivan2012overview}. 

G-PCC adopts a different strategy, introducing two important geometry encoding modes: octree-based G-PCC and trisoup-based G-PCC. Octree-based G-PCC begins by quantizing the point cloud and optionally merging points with identical locations. The quantized point cloud is then represented using an octree in the 3D space, allowing for efficient point cloud representation of varying densities. The octree structure is encoded using context-based arithmetic coding, with accompanying recording of occupancy information for each octant. Trisoup-based G-PCC serves as a powerful complement to the octree decomposition, in which the occupied leaf nodes correspond to 3D cubes that may contain multiple points. Each occupied 3D cube at the leaf level is represented by surfaces composed of triangle strips, connecting vertices along the edges of the 3D cube. Rather than encoding the point coordinates, the information about these triangles is encoded. 

V-PCC has proven effective in compressing solid point clouds but is less suited for sparse point clouds. Furthermore, it exhibits a higher encoding time complexity when contrasted with G-PCC. In our current work, we strive to bridge the performance gap between G-PCC and V-PCC for solid point clouds by improving octree-based G-PCC with a hierarchical prior while inheriting its computational efficiency. A concurrent work named ``Improved Trisoup''~\cite{mpeg:m59973} also shows impressive gains.

\noindent\textbf{Deep learning-based PCC}.
As a binary signal on a voxel grid,
point cloud geometry is amenable to compression by Convolutional Neural Networks (CNNs)~\cite{quach2019learning,quach2022survey}. However, the computational complexity of standard convolution over the entire voxel grid can be substantial. Researchers have explored block partitioning and sparse convolution to tackle this issue~\cite{quach2020improved,wang2021lossy}. Lazzarotto~\etal ~\cite{lazzarotto2021block,lazzarotto2021learning} applied residual connection and block prediction for learning-based PCC. Empirical evidence shows that deep learning-based PCC systematically overfits the point cloud densities in the training set~\cite{guarda2021adaptive}. Guarda~\etal~\cite{guarda2021adaptive} instead trained multiple CNNs for different point cloud densities. During compression, the optimal CNN is selected for each point cloud block, and its corresponding index is recorded as side information. Another interesting learning-based PCC method is PCGCv2~\cite{wang2021multiscale}, which presents a multi-scale learning scheme for reconstructing point cloud geometry through progressive resampling. 

Deep learning-based lossless compression of point clouds~\cite{fu2022octattention} can be extended to lossy compression by incorporating a downsampling step. Nguyen~\etal~\cite{nguyen2021learning} introduced a CNN with masked convolutions for lossless coding of point cloud geometry. They initially implemented a sequential context-based coding scheme, which is rather slow, and later accelerated it by estimating some occupancy probabilities in parallel~\cite{nguyen2021multiscale}. Such sequential dependencies were entirely removed in~\cite{que2021voxelcontext} by predicting the voxel occupancy using the parent-level information.

While deep learning-based PCC exhibits impressive rate-distortion performance, it has notable drawbacks in terms of time complexity, scalability to large-scale point clouds, and generalization across point clouds of different densities.

\subsection{Point Cloud Geometry Super Resolution}\label{subsec:pcsr}
Before delving into point cloud geometry super resolution or upsampling, it is essential to establish a clear understanding of the downsampling process, which represents the inverse operation. Point cloud geometry downsampling can be achieved through set downsampling and grid downsampling. Set downsampling decimates points in the original set without changing the voxel resolution, while grid downsampling changes the number of points by revoxelizing the point cloud (\ie, changing the volumetric resolution). Set downsampling excels at preserving the overall geometry and finer details of the point cloud but may introduce a less regular point distribution. In contrast, grid downsampling reduces the volumetric resolution, making it suitable for compact representation.

\begin{figure*}[!t]
\centering
\begin{minipage}[b]{\linewidth}
\begin{center}
\resizebox{.8\linewidth}{!}{\hpsrpccEncoder}
\end{center}
\footnotesize{(a) Encoder. The original point cloud $\V$ undergoes a series of downsampling operations, resulting in a point cloud pyramid $\{\V^{(k)}\}_{k=0}^{K}$, where each level $k$ is downsampled by a factor of $s_k$ for $k=0,\cdots,K$. Subsequently, we construct the hierarchical prior $\{\bsigma^{(k)}\}_{k=1}^{K}$ based on the point cloud pyramid. To encode $\V$, the base point cloud $\V^{(K)}$ and the hierarchical prior $\{\bsigma^{(k)}\}_{k=1}^K$ are both subjected to lossless encoding.} 
\vspace{3mm}
\end{minipage}
\begin{minipage}[b]{\linewidth}
\begin{center}
\resizebox{.7\linewidth}{!}{\hpsrpccDecoder}
\end{center} 
\footnotesize{(b) Decoder. The received bitstreams are decoded in a lossless manner, resulting in the reconstructed base point cloud $\V^{(K)}$ and the hierarchical prior $\{\bsigma^{(k)}\}_{k=1}^K$. We then progressively super-resolve the base point cloud $\V^{(K)}$ to generate the final reconstructed point cloud $\hV$ with the hierarchical prior.} 
\end{minipage}
\caption{System diagram of the proposed hierarchical prior-based super resolution for point cloud geometry compression.}\label{fig:framework}
\end{figure*}

Currently, most point cloud geometry super resolution methods~\cite{alexa2003computing,huang2013edge,dinesh2022point,yu2018pu,qian2021deep,liu2022spu,liu2022pufa} are designed for set downsampling. Nevertheless, grid downsampling is considered in octree-based G-PCC, which expects distinct post-processing super resolution methods~\cite{akhtar2020point,akhtar2022pu,fan2022deep,garcia2020geometry,borges2022fractional}. Akhtar~\etal~\cite{akhtar2020point,akhtar2022pu} predicted the occupancy of child points in the decoded point cloud by a neural network. Building upon~\cite{akhtar2020point}, Fan~\etal~\cite{fan2022deep} proposed a single model, capable of enhancing decoded point clouds with varying degrees of distortions. While these deep learning-based techniques yield noticeable improvements, they come with added computational complexity, limiting their wide adoption in time-sensitive applications. Garcia~\etal~\cite{garcia2020geometry} proposed a neighborhood inheritance-based super resolution method for dynamic point clouds, which constructs a dictionary of child nodes based on the neighborhood configuration from previous frames. Borges~\etal~\cite{borges2022fractional} proposed SRLUT, making several improvements over~\cite{garcia2020geometry}. One notable enhancement is the extension of fractional resampling capability. Additionally, SRLUT is an intra super resolution method with improved practicability. 

Although post-processing techniques such as SRLUT can reduce compression artifacts without increasing the bitrate, their rate-distortion performance is often sub-optimal. This work introduces a hierarchical prior-based super resolution method trading off the rate and distortion. Although the proposed HPSR-PCGC and SRLUT~\cite{borges2022fractional} share some similarities, \eg, employing neighborhood-based point clustering and frequency-based occupancy estimation to construct interpolation patterns, they differ in substantial ways. First, as a post-processing method, SRLUT can not trade off the rate and distortion, while HPSR-PCGC presents a principled approach of doing so by adjusting the hyperparameters during the hierarchical prior construction. Second, the assumption of the cross-scale self-similarity by SRLUT is not always valid, particularly for sparse point clouds. HPSR-PCGC relaxes the assumption by constructing interpolation patterns at the encoder side. Although sending these interpolation patterns requires more bits, our results have demonstrated that such a design is worthwhile. Third, decoding complexity is generally considered more important than encoding complexity~\cite{dupont2021coin}. SRLUT constructs interpolation patterns at the decoder side, whereas our method does so at the encoder side, prioritizing decoding complexity over encoding complexity. Lastly, SRLUT relies heavily on data augmentation to refine a finer geometry using a coarser geometry. In contrast, HPSR-PCGC constructs interpolation patterns within the same scale, and requires no data augmentation.

\section{Proposed HPSR-PCGC}\label{sec:method}
This section details our improved point cloud geometry compression method, HPSR-PCGC, through hierarchical prior-based super resolution. The system diagram is illustrated in Fig.~\ref{fig:framework}, which includes modules of successive downsampling and hierarchical prior construction in the encoder and hierarchical prior-based super resolution in the decoder.

\subsection{Successive Downsampling}
\begin{figure}[!t]
\centering
\includegraphics[width=.55\linewidth]{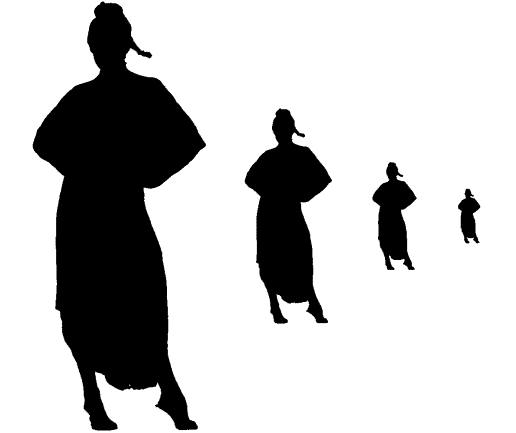}
\caption{Illustration of a point cloud pyramid produced by the successive downsampling. $\V, \V^{(0)}, \V^{(1)}$, and $\V^{(2)}$ are shown from left to right.}\label{fig:pyramid}
\end{figure}

The successive downsampling module is crucial in generating the necessary information for constructing the hierarchical prior. It accepts as input the original point cloud $\V\in\mathbb{R}^{N\times 3}$ and the downsampling factor $q\in(0,1)$, where a smaller $q$ indicates a coarser-grained (\ie, heavier) downsampling level. The module produces a sequence of point clouds $\{\V^{(k)}\}_{k=0}^{K}$, which we refer to as a point cloud pyramid by analogy to image pyramid~\cite{adelson1984pyramid} in signal processing:
\begin{align}
    \V^{(0)}&=\mathrm{unique}\left(\left[\V/2^{L+1-K}\right]\right),\\
    \V^{(k)}&=\mathrm{unique}\left(\left[\V^{(k-1)}/2\right]\right)\ \mbox{for $k=1,\cdots,K-1$,}\\
    \V^{(K)}&=\mathrm{unique}\left(\left[\V^{(K-1)}\times 2^L\times q\right]\right).
\end{align}
$\mathrm{unique}(\cdot)$ is the function to remove duplicated points, $[\cdot]$ indicates the rounding function, and $L = \lceil\log_2(1/q)\rceil-1$ relates to the maximum level of downsampling by $K \le L+1$. Fig.~\ref{fig:pyramid} visually illustrates a point cloud pyramid with $K=2$ and $q=1/8$.

\begin{figure*}[!t]
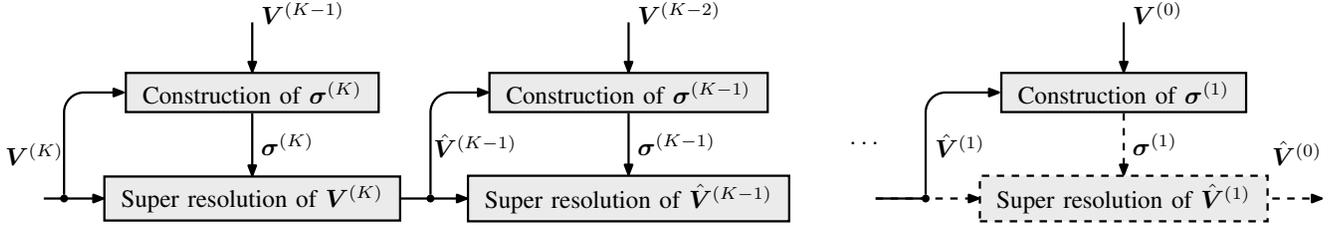

\centering
\resizebox{.99\linewidth}{!}{
\hpcPipeline
}
\caption{Pipeline of the hierarchical prior construction.}\label{fig:HPC}
\end{figure*}

\begin{figure*}[!t]
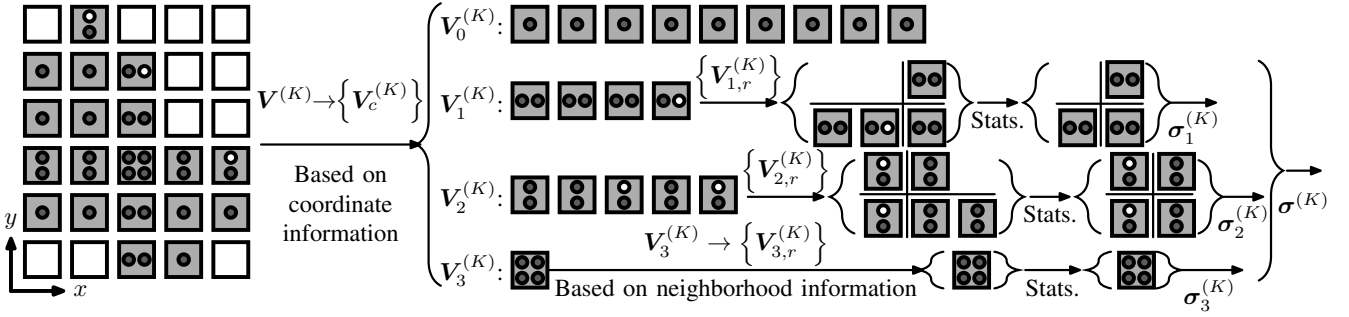

\centering
\resizebox{.99\linewidth}{!}{\demo}
\caption{2D illustration for constructing the interpolation patterns $\bsigma^{(K)}$ that help map $\V^{(K)}$ to an approximation of $\V^{(K-1)}$, where $2^L\times q=3/4$ and the neighborhood consists of the left and right voxels only. Gray/white squares indicate occupied/void voxels in $\V^{(K)}$, while gray/white circles indicate occupied/void voxels in $\V^{(K-1)}$. With a factor of $3/4$, points denoted by gray circles are downsampled to the same point denoted by the circumscribed square. ``Stats.'' indicates simple frequency-based statistical analysis.}\label{fig:hpK}
\end{figure*}

\begin{figure*}[!t]
\centering
\resizebox{.7\linewidth}{!}{
\anotherdemo
}
\caption{2D illustration for constructing the interpolation patterns $\bsigma^{(k)}$ where $k=K-1,\cdots,1$ that help map $\hV^{(k)}$ to an approximation of $\V^{(k-1)}$. Only left and right neighbors are considered. Gray/white squares indicate occupied/void voxels in $\hV^{(k)}$, and gray/white circles indicate occupied/void voxels in $\V^{(k-1)}$. When $\V^{(k-1)}$ is downsampled with a factor of $1/2$, points denoted by gray circles are merged to the same point denoted by the circumscribed square. We first partition $\hV^{(k)}$ into several clusters $\{\hV^{(k)}_{r}\}$ based on neighborhood information, and then obtain the prior $\bsigma^{(k)}$ based on frequency statistics.}\label{fig:hpk}
\end{figure*}

\subsection{Hierarchical Prior Construction}\label{sec:hpc}
Directly upscaling $\V^{(K)}$ without interpolation may lead to severe distortions. To address this issue, we construct a hierarchical prior that facilitates the coarse-to-fine super resolution of $\V^{(K)}$ during decoding. Achieving a lossless reconstruction of $\V$ would require full prior knowledge on how each point in $\V^{(k)}$ should be interpolated, either progressively towards $\V^{(k-1)}$ or in a single step towards $\V$. We refer to this knowledge as the \textit{interpolation pattern}, as will be immediately clear. However, this will cost superabundant bits to encode such prior information, perhaps even more bits than direct lossless compression of $\V$. Alternatively, interpolating all points uniformly using a single pattern would often be ineffective. A more approachable way is to first perform point clustering and then design an interpolation pattern for all points in one cluster, where we have good control of the clustering to trade off the rate and distortion.

As shown in Fig.~\ref{fig:HPC}, the proposed hierarchical prior consists of $K$ sets of interpolation patterns $\{\bsigma^{(k)}\}_{k=1}^K$ that allow progressively mapping $\V^{(K)}$ to an approximation of $\V$. We leverage two types of information available in the decoder to perform point clustering: voxel coordinates and local neighbors. The incorporation of coordinate information gives a special treatment of non-uniform grid downsampling to obtain $V^{(K)}$ when $2^L\times q>1/2$. The utilization of neighborhood information is rooted in the assumption of non-local geometric similarity. By conducting the same point clustering process at the decoder side, we can interpolate the base point cloud using the transmitted interpolation patterns.

To have an intuitive understanding, Fig.~\ref{fig:hpK} illustrates the construction of interpolation patterns $\bsigma^{(K)}$ for mapping $\V^{(K)}$ to an approximation of $\V^{(K-1)}$ using a 2D example. We assume that $2^L\times q=3/4$, and the neighborhood consists of only the left and right voxels. To begin with, we partition $\V^{(K)}$ into four parts, denoted as ${\V^{(K)}_c}$, where $c\in\{0,1,2,3\}$, based on the coordinate information. Points in $\V^{(K)}_0$ have a unique correspondence; points in $\V^{(K)}_1$ and $\V^{(K)}_2$ have the one-to-two correspondence only in $x$-axis and $y$-axis, respectively; and points in $\V^{(K)}_3$ have one-to-two correspondences in both axes. Next, for $\V^{(K)}_c$ where $c>0$,  we further divide them into clusters, denoted as ${\V^{(K)}_{c,r}}$, based on neighborhood information. For instance, $\V^{(K)}_{2}$ is divided into four clusters: the upper-left cluster, $\V^{(K)}_{2,0}$, with both left and right voxels being void, the upper-right cluster, $\V^{(K)}_{2,1}$, with only the left voxel being void, the lower-left cluster, $\V^{(K)}_{2,2}$, with only the right voxel being void, and the lower-right cluster, $\V^{(K)}_{2,3}$, with both left and right voxels being occupied. Finally, for $\V^{(K)}_c$ where $c>0$, we obtain the interpolation pattern in the form of $\bsigma^{(K)}_c$ based on simple frequency-based statistics.

\noindent\textbf{Construction of base point cloud priors $\bsigma^{(K)}$}.
In line with SRLUT~\cite{borges2022fractional}, when $1/2<2^L\times q<1$, non-uniform downsampling occurs, leading to one-to-one and one-to-two correspondences between voxels after and before downsampling along each coordinate axis. This results in eight distinct cases. Consequently, we initially divide $\V^{(K)}$ into eight clusters, denoted as $\{\V^{(K)}_c\}_{c=0}^7$, based on 3D voxel coordinates. Cluster ${\V}^{(K)}_0$ contains points that have a unique correspondence with points in ${\V}^{(K-1)}$, allowing for perfect reconstruction. Points in ${\V}^{(K)}_1$, ${\V}^{(K)}_2$, and ${\V}^{(K)}_4$ have the one-to-two correspondence only along the $x$-axis, $y$-axis, and $z$-axis, respectively. Points in ${\V}^{(K)}_3$, ${\V}^{(K)}_5$, and ${\V}^{(K)}_6$ have the one-to-one correspondence only along the $z$-axis, $y$-axis, and $x$-axis, respectively. Points in ${\V}^{(K)}_7$ have one-to-two correspondences along all three coordinate axes, resulting in the worst case of one-to-eight correspondence. Then, we have a point in ${\V}^{(K)}_c$ resulting from at most $M_c$ points in ${\V}^{(K-1)}$, where
\begin{equation}\label{eq:T}
    M_c=\begin{cases}
    1& \text{if}\ c=0,\\
    2& \text{if}\ c=1, 2\ \text{or}\ 4,\\
    4& \text{if}\ c=3, 5\ \text{or}\ 6,\\
    8& \text{if}\ c=7.
    \end{cases}
\end{equation}
To derive a more accurate prior, we further partition the clusters $\{\V^{(K)}_c\}_{c=1}^7$ based on local neighborhood information. Specifically, we define $\mathcal{N}_K = \{(x_n, y_n, z_n)\}$ as the set of neighboring voxels of $(x,y,z) \in \V^{(K)}$. We encode the occupancy of each neighbor using one bit and summarize this information using an integer value:
\begin{equation}\label{eq:neighborhood}
    \phi_K(x,y,z) = \sum_{n=0}^{\vert \mathcal{N}_K
\vert-1}\left(\mathbb{I}\left[(x_{n},y_{n},z_{n})\in \V^{(K)}\right]\right)\times 2^n,
\end{equation}
where $\mathbb{I}[\cdot]$ represents the indicator function. This encoding allows the local neighborhood information to be captured in a compact form. Assuming that $\mathcal{R}^{(K)}_{c}=\{\phi_K(x,y,z)\}_{(x,y,z)\in{\V}^{(K)}_{c}}$ represents the set of the observed neighborhood information for cluster ${\V}^{(K)}_{c}$, we further partition ${\V}^{(K)}_{c}$ into $\vert\mathcal{R}^{(K)}_{c}\vert$ different subsets $\V^{(K)}_{c,r}$:
\begin{equation}\label{eq:Vdrc}
    \V^{(K)}_{c,r}=\left\{(x,y,z)\in\V^{(K)}_{c}|\phi_K(x,y,z)=r\right\},\ \text{for}\ r\in \mathcal{R}^{(K)}_{c}.
\end{equation}
That is, points in ${\V}^{(K)}_{c}$ that share the same local patterns form finer partitions, which facilitates more accurate modeling of the dependencies and characteristics of the point cloud, and leads to improved accuracy in hierarchical prior construction and decoder-side super resolution.

To construct the interpolation pattern for $\V^{(K)}_{c,r}$, we begin by denoting all possible corresponding points of $(x,y,z) \in \V^{(K)}_{c,r}$ as $\mathcal{C}_{K,c}=\{(x_{m}, y_{m}, z_{m})\}_{m=0}^{M_c-1}$, which represent candidate interpolation points. Next, we compute the occurrence number of the $m$-th point, denoted as $p^{(K)}_m$, in $\V^{(K-1)}$ for points in $\V^{(K)}_{c,r}$:
\begin{equation}
    p^{(K)}_m = \sum_{(x,y,z)\in \V^{(K)}_{c,r}} \mathbb{I}\left[(x_{m},y_{m},z_{m})\in \V^{(K-1)}\right].
\end{equation}
The above indicator function returns $1$ if the $m$-th point is found in $\V^{(K-1)}$ and $0$ otherwise. We convert $p^{(K)}_m$ into a frequency $f^{(K)}_{m}$ by dividing $p^{(K)}_m$ by the total number of points in $\V^{(K)}_{c,r}$. This frequency represents the likelihood of the $m$-th point being occupied. If the frequency $f^{(K)}_{m}$ is greater than or equal to $0.5$, we interpolate this point. Otherwise, we leave it empty. Finally, we define the interpolation pattern, $\sigma^{(K)}_{c,r}$, for points in $\V^{(K)}_{c,r}$, based on frequency-based statistics:
\begin{equation}\label{eq:imp}
    \sigma^{(K)}_{c,r}=\sum_{m=0}^{M_c-1}\mathbb{I}\left[f^{(K)}_{m}\ge 0.5\right]\times 2^m.
\end{equation}

\noindent\textbf{Construction of intermediate priors $\bsigma^{(k)}$ for $k=K-1,\cdots,1$}. 
To construct the interpolation pattern $\bsigma^{(k)}$, we may have to obtain the reconstructed $\hat{\V}^{(k)}$ first because the original $\V^{(k)}$ may not be available or cannot be perfectly reconstructed at the decoder side. It is important to note that in constructing $\bsigma^{(k)}$, we do not require coordinate information for point clustering, as all points in $\hV^{(k)}$ have one-to-two correspondences in all axes, with a downsampling factor of $1/2$. Fig.~\ref{fig:hpk} illustrates a 2D example for the construction of the interpolation pattern $\bsigma^{(k)}$ that helps map $\hV^{(k)}$ to $\hV^{(k-1)}$ as an approximation of $\V^{(k-1)}$.

\subsection{Hierarchical Prior-based Super Resolution}\label{sec:hpsr}
\begin{figure}[!t]
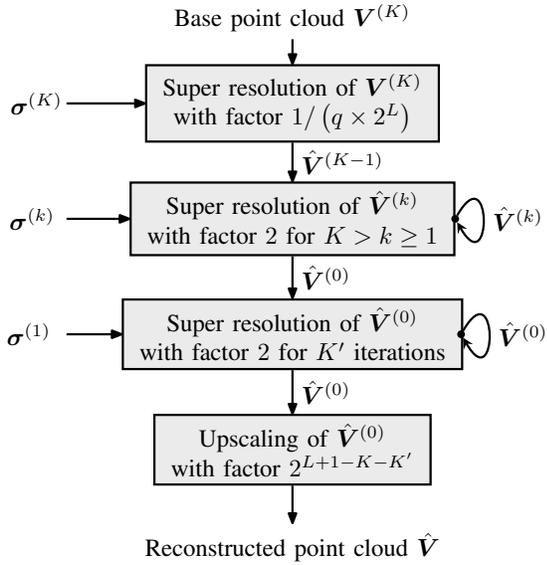

\centering
\resizebox{.85\linewidth}{!}{\hpsr}
\caption{Pipeline of our hierarchical prior-based super resolution, where $K'$ is a user-defined parameter and $0 \leq K' \leq L+1-K$.}\label{fig:hpsr}
\end{figure}

The pipeline of our hierarchical prior-based super resolution module is illustrated in Fig.~\ref{fig:hpsr}. The primary objective is to super-resolve the base point cloud progressively to improve the reconstruction quality. After that, the super-resolved point cloud is upscaled to match the scale of the original data.

\begin{figure*}[!t]
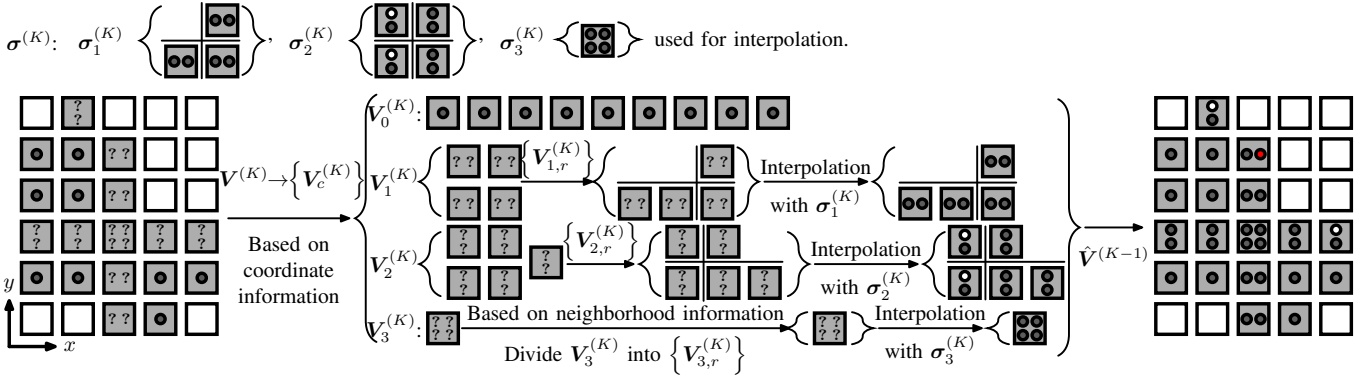

\centering
\resizebox{\linewidth}{!}{\idemo}
\caption{2D illustration for interpolating the base point cloud $\V^{(K)}$ with $\bsigma^{(K)}$ constructed in Fig.~\ref{fig:hpK}. The question mark ``?'' stands for our ignorance of the voxel occupancy when decoding. We partition $\V^{(K)}$ into $\{\V^{(K)}_{c,r}\}$ based on coordinate and neighborhood information (see Subsec.~\ref{sec:hpc}). We interpolate points in $\V^{(K)}_0$ by direct upscaling. For points in $\V^{(K)}_c$ where $c>0$, we interpolate them with $\bsigma^{(K)}_c$, giving rise to $\hV^{(K-1)}$. The red circle indicates the extra added point in $\hV^{(K-1)}$ compared to $\V^{(K-1)}$.}\label{fig:interpolationK}
\end{figure*}

\noindent{\textbf{Super resolution of} $\V^{(K)}$}. 
The same method described in Subsec.~\ref{sec:hpc} is utilized to partition $\V^{(K)}$. In detail, based on the coordinate information, $\V^{(K)}$ is divided into eight subsets $\V^{(K)}_{c}$, where $c\in\{0,\cdots,7\}$. Further division into $\{\V^{(K)}_{c,r}\}$ based on neighborhood information is performed. According to the decoded prior $\bsigma^{(K)}$, points in $\V^{(K)}_{0}$ is directly upscaled by dividing them with a factor of $q\times 2^L$, and $\hV^{(K-1)}$ is initialized as $\left[{\V^{(K)}_{0}}/\left(q\times 2^L\right)\right]$. Each $\V^{(K)}_{c,r}$ contributes points with the interpolation pattern $\sigma^{(K)}_{c,r}$ to $\hV^{(K-1)}$. We repeat this process for all subsets $\{\V^{(K)}_{c,r}\}$ to obtain the interpolated point cloud $\hV^{(K-1)}$. Fig.~\ref{fig:interpolationK} provides a 2D illustration for better comprehension.

\begin{figure*}[!t]
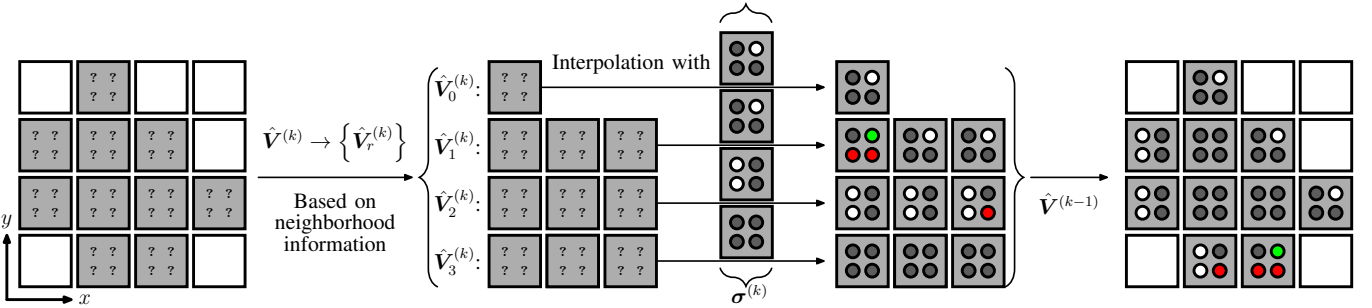

\centering
\resizebox{\linewidth}{!}{\anotheridemo}
\caption{2D illustration for interpolating $\hV^{(k)}$ to $\hV^{(k-1)}$ with $\bsigma^{(k)}$. The question mark ``?'' stands for our ignorance of the voxel occupancy when decoding. We partition $\hV^{(k)}$ into $\{\hV^{(k)}_{r}\}$ based on neighborhood information. For a given $r$, we interpolate points in $\hV^{(k)}_r$ with $\sigma^{(k)}_r\in\bsigma^{(k)}$, resulting in the interpolated point cloud $\hV^{(k-1)}$. The red/green circle indicates the extra added/removed point in $\hV^{(k-1)}$ compared to $\V^{(k-1)}$.}\label{fig:interpolationk}
\end{figure*}

\noindent{\textbf{Super resolution of $\hV^{(k)}$ for $k=K-1,\cdots, 1$}}. 
The process continues with the partitioning of $\hV^{(k)}$ to $\{\hV^{(k)}_{r}\}$ based on neighborhood information. Initially, $\hV^{(k-1)}$ is an empty set, and an interpolation process is applied to all points in $\hV^{(k)}_{r}$ using $\sigma^{(k)}_{r}$. $\hV^{(k-1)}$ is derived when all subsets $\{\hV^{(k)}_{r}\}$ are processed. A 2D illustration is provided in Fig.~\ref{fig:interpolationk}. This procedure is repeated until all $K-1$ scales are exhausted to reach $\hV^{(0)}$.

\noindent{\textbf{Super resolution of} $\hV^{(0)}$}. 
As validated by SRLUT~\cite{borges2022fractional}, point cloud geometry exhibits strong self-similarity across scales, particularly for solid point clouds. At low bitrates, the super-resolved point cloud $\hV^{(0)}$ may be solid, which is amenable to further super resolution by reusing the last interpolation pattern $\bsigma^{(1)}$ for $K'$ iterations, where $0 \leq K' \leq L+1-K$. In particular, the set of the observed neighborhood information for $\hV^{(1)}$, denoted as $\mathcal{R}^{(1)}=\{\phi_1(x,y,z)\}_{(x,y,z)\in{\hV}^{(1)}}$, is adopted to partition $\hV^{(0)}$. This partitioning results in $\vert\mathcal{R}^{(1)}\vert+1$ subsets, with the additional subset accommodating points with neighborhood information not found in $\mathcal{R}^{(1)}$. Points within this extra subset undergo direct upscaling, whereas the remaining points are interpolated using $\bsigma^{(1)}$.

\noindent{\textbf{Upscaling of} $\hV^{(0)}$}. 
The final reconstructed point cloud, denoted as $\hV$, is obtained by upscaling $\hV^{(0)}$ to match the scale of the original point cloud $\V$: $\hV=\left[\hV^{(0)}/{2^{K'+K-L-1}}\right]$.

\begin{table*}[!t]
\centering
\caption{D1- and D2-BDBR savings of the proposed methods (HPSR-PCGC \& HPSR-PCGC-RDO) against G-PCC (octree) and G-PCC (trisoup)} \label{tab:bdbr}
\begin{tabular}{lcccccc}
  \toprule
  \multirow{2}{*}{Point Cloud} & \multicolumn{2}{c}{HPSR-PCGC vs. G-PCC (octree)} & \multicolumn{2}{c}{HPSR-PCGC vs. G-PCC (trisoup)} & \multicolumn{2}{c}{HPSR-PCGC-RDO vs. G-PCC (trisoup)} \\
  \cmidrule(lr){2-3}
  \cmidrule(lr){4-5}
  \cmidrule(lr){6-7}
  & D1 & D2 & D1 & D2 & D1 & D2 \\
  \midrule
  basketball\_player\_vox11\_00000200~\cite{mpeg:m42816} & $-79.2\%$ & $-68.0\%$ & $-56.7\%$ & $-38.9\%$ & $-64.6\%$ & $-50.1\%$ \\
  dancer\_vox11\_00000001~\cite{mpeg:m42816} & $-77.0\%$ & $-63.9\%$ & $-53.7\%$ & $-32.1\%$ & $-62.9\%$ & $-45.1\%$ \\
  facade\_00064\_vox11~\cite{mpeg:m38678} & $-78.0\%$ & $-60.0\%$ & $-68.2\%$ & $-48.4\%$ & $-77.1\%$ & $-60.0\%$ \\
  longdress\_vox10\_1300~\cite{mpeg:m40059} & $-73.3\%$ & $-58.4\%$ & $-32.8\%$ & $-19.8\%$ & $-57.0\%$ & $-49.9\%$\\
  loot\_vox10\_1200~\cite{mpeg:m40059} & $-75.0\%$ & $-58.4\%$ & $-39.9\%$ & $-10.8\%$ & $-59.3\%$ & $-41.2\%$\\
  queen\_0200~\cite{mpeg:m40050} & $-75.0\%$ & $-59.4\%$ & $-34.3\%$ & $-16.7\%$ & $-59.6\%$ & $-49.0\%$\\
  redandblack\_vox10\_1550~\cite{mpeg:m40059} & $-69.9\%$ & $-53.2\%$ & $-32.1\%$ & $-7.6\%$ & $-59.1\%$ & $-46.5\%$ \\
  soldier\_vox10\_0690~\cite{mpeg:m40059} & $-73.8\%$ & $-59.9\%$ & $-30.9\%$ & $-19.8\%$ & $-53.5\%$ & $-45.1\%$ \\
  thaidancer\_viewdep\_vox12~\cite{mpeg:m42914} & $-66.9\%$ & $-53.8\%$ & $-35.9\%$ & $-13.6\%$ & $-57.2\%$ & $-36.9\%$ \\
  \midrule
  \textbf{Solid (Average)} & $-74.3\%$ & $-59.4\%$ & $-42.7\%$ & $-23.1\%$ & $-61.1\%$ & $-47.1\%$ \\
  \midrule
  boxer\_viewdep\_vox12~\cite{mpeg:m42914} & $-72.5\%$ & $-63.3\%$ & $-29.5\%$ & $-10.8\%$ & $-36.0\%$ & $-22.9\%$ \\
  facade\_00009\_vox12~\cite{mpeg:m38678} & $-48.3\%$ & $-24.3\%$ & $-56.3\%$ & $3.7\%$ & $-67.8\%$ & $-11.1\%$ \\
  facade\_00015\_vox14~\cite{mpeg:m38678} & $-68.8\%$ & $-55.2\%$ & $-81.2\%$ & $-50.3\%$& $-82.7\%$ & $-39.7\%$ \\
  frog\_00067\_vox12~\cite{mpeg:m38678} & $-60.9\%$ & $-54.6\%$ & $-63.1\%$ & $-17.9\%$ & $-71.0\%$ & $-4.0\%$ \\
  head\_00039\_vox12~\cite{mpeg:m38678} & $-73.0\%$ & $-67.5\%$ & $-83.3\%$ & $-58.7\%$ & $-84.6\%$ & $-53.9\%$ \\
  house\_without\_roof\_00057\_vox12~\cite{mpeg:m38678} & $-76.3\%$ & $-60.3\%$ & $-77.0\%$ & $-30.2\%$ & $-80.2\%$ & $-17.1\%$ \\
  longdress\_viewdep\_vox12~\cite{mpeg:m42914} & $-66.2\%$ & $-55.7\%$ & $-12.3\%$ & $-20.2\%$ & $-30.5\%$ & $-34.5\%$ \\
  loot\_viewdep\_vox12~\cite{mpeg:m42914} & $-68.5\%$ & $-59.5\%$ & $-24.2\%$ & $-7.9\%$ & $-39.4\%$ & $-22.4\%$ \\
  redandblack\_viewdep\_vox12~\cite{mpeg:m42914} & $-60.9\%$ & $-51.0\%$ & $-7.6\%$ & $-7.2\%$ & $-32.3\%$ & $-26.5\%$ \\ 
  soldier\_viewdep\_vox12~\cite{mpeg:m42914} & $-66.5\%$ & $-58.6\%$ & $-16.2\%$ & $-17.5\%$ & $-32.9\%$ & $-30.2\%$ \\
  \midrule
  \textbf{Dense (Average)} & $-66.2\%$ & $-55.0\%$ & $-45.1\%$ & $-21.7\%$ & $-55.7\%$ & $-26.2\%$ \\
  \midrule
  egyptian\_mask\_vox12~\cite{mpeg:m38678} & $-14.0\%$ & $8.4\%$ & $-17.4\%$ & $82.2\%$ & $-39.9\%$ & $30.9\%$ \\
  shiva\_00035\_vox12~\cite{mpeg:m38678} & $-38.5\%$ & $-5.8\%$ & $-47.0\%$ & $-0.1\%$ & $-59.1\%$ & $-46.5\%$ \\
  ulb\_unicorn\_vox13~\cite{mpeg:m41742} & $-25.9\%$ & $27.4\%$ & $-95.5\%$ & $-47.4\%$ & $-96.3\%$ & $-49.7\%$ \\
  \midrule
  \textbf{Sparse (Average)} & $-26.1\%$ & $10.0\%$ & $-53.3\%$ & $11.6\%$ & $-68.3\%$ & $-19.3\%$ \\
  \midrule
  \textbf{All (Cat1A Average)} & $-64.0\%$ & $-48.0\%$ & $-45.2\%$ & $-17.7\%$ & $-59.7\%$ & $-33.8\%$ \\
  \bottomrule
\end{tabular}
\end{table*}

\subsection{Base and Prior Coders}
In our implementation, we employ the octree-based G-PCC as the base encoder/decoder for the base point cloud $\V^{(K)}$. The hierarchical prior, represented as integer values, can be directly written/read in the prior encoder/decoder. 

\section{Experiments}\label{sec:experiment}
\begin{figure}[!t]
\centering
\includegraphics[width=\linewidth]{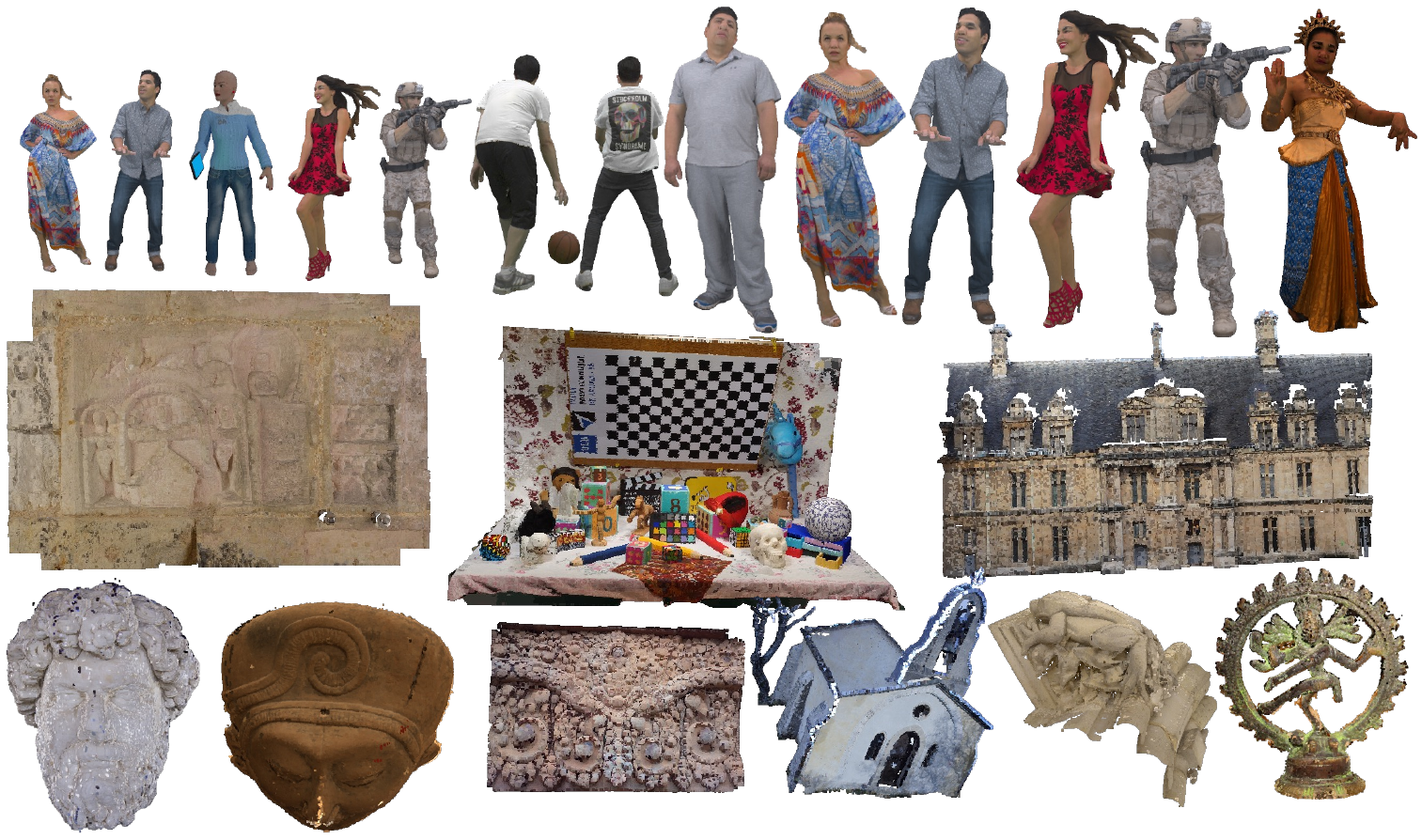}
\caption{Visualization of $22$ point clouds in the MPEG Cat1A dataset, where four point clouds have two versions: $10$ bits and $12$ bits.}\label{fig:mpegcat1a}
\end{figure}

In this section, we conducted experiments to evaluate the proposed HPSR-PCGC under the C2 condition (lossy geometry and lossy attributes) by following the Common Test Conditions (CTC) for G-PCC~\cite{mpeg2021ctc}. The experiments were carried out on the MPEG-Cat1A dataset, which consists of $22$ point clouds~\cite{mpeg:m42816,mpeg:m42914,mpeg:m38678,mpeg:m40050,mpeg:m40059,mpeg:m41742}. These point clouds are categorized as ``solid'' (nine point clouds), ``dense'' (ten point clouds), and ``sparse'' (three point clouds), based on the categorization used in the PCC community~\cite{new-ctc}. Snapshots of these $22$ point clouds (with color attributes) are provided in Fig.~\ref{fig:mpegcat1a}. To measure the distortion, we considered point-to-point (D1) and point-to-plane (D2) distance metrics~\cite{tian2017geometric}. To evaluate rate-distortion performance gains, we reported the Bj{\o}ntegaard-Delta BitRate (BDBR)~\cite{VCEG-M33}. 

To cover a large Peak Signal-to-Noise Ratio (PSNR) range, we related the geometry quantization/downsampling parameter $s$ suggested by MPEG G-PCC (octree) to 
$q$ in the proposed HPSR-PCGC by 
\begin{equation}
    q = f\left(f(s)\right),
\end{equation}
where
\begin{equation}
    f(s)=\begin{cases}
    \frac{a-1}{b} & \text{if}\ s=\frac{a}{b}>0.5,\ \mbox{and $a$ and $b$ are coprime}\\
    s/2 & \text{otherwise}.
    \end{cases}
\end{equation}
The neighboring set $\mathcal{N}_K$ contains eighteen voxels that share a line or face with the center point, while $\mathcal{N}_k$ for $k<K$ contains six voxels that share a face with the center point. The default hyperparameters of HPSR-PCGC are set as follows: $K = \min(L+1, 2)$ and $K' = \min(2, L+1-K)$, where $L = \lceil\log_2(1/q)\rceil - 1$. The implementation of HPSR-PCGC can be found at \url{https://github.com/lidq92/mpeg-pcc-tmc13/tree/hpsr_pcgc}.

\subsection{BDBR Comparison to G-PCC}
Our HPSR-PCGC utilizes the lossless G-PCC (octree) as the base encoder/decoder, supplemented with pre- and post-processing modules. 
We compare HPSR-PCGC with G-PCC using its latest available software, MPEG-PCC-TMC13 v14.0~\cite{GPCCv14} as the anchor method. 
The left part of Table~\ref{tab:bdbr} presents the BDBR savings for HPSR-PCGC compared to G-PCC (octree and trisoup). For D1-BDBR, HPSR-PCGC achieves more significant savings for solid point clouds ($74.3\%$) than sparse point clouds ($26.1\%$), relative to G-PCC (octree). This is expected because our assumption of non-local geometry similarity becomes less valid as the point cloud density decreases. When comparing HPSR-PCGC to G-PCC (trisoup), fewer performance variations for point clouds with varying densities are observed, which is reasonable since both HPSR-PCGC and G-PCC (trisoup) are more effective in handling denser point clouds. The performance variations within the same density category may be attributed to the content variations presented in the dataset.

Regarding D2-BDBR, HPSR-PCGC exhibits reduced savings compared to D1-BDBR savings for solid and dense point clouds. HPSR-PCGC even shows a D2-BDBR overhead on sparse point clouds. This discrepancy reveals that HPSR-PCGC interpolates points without imposing geometric constraints on the surface. Consequently, the interpolated points may not align perfectly with the underlying surface of the point cloud, causing large point-to-plane (D2) errors.

\begin{figure}[!t]
    \centering
    \includegraphics[width=.8\linewidth]{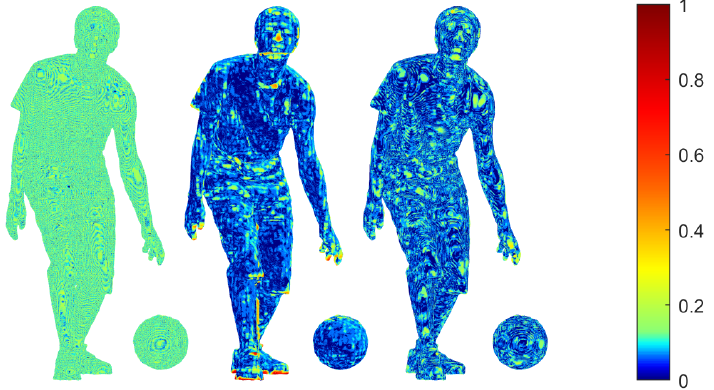}
    \caption{Error maps of ``basketball\_player\_vox11\_00000200''. Left: G-PCC (octree), bpp = 0.07, D1-PSNR = 64.46. Middle: G-PCC (trisoup), bpp = 0.05, D1-PSNR = 67.96. Right: HPSR-PCGC, bpp = 0.03, D1-PSNR = 70.29.}\label{fig:solid-error}
\end{figure}

\begin{figure}[!t]
    \centering
    \includegraphics[width=.55\linewidth]{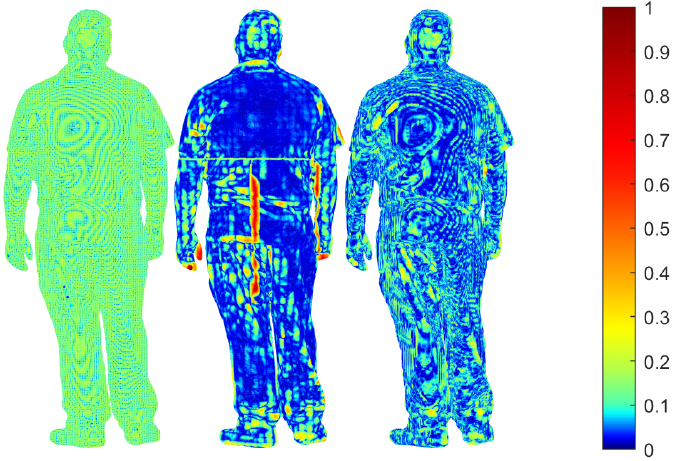}
    \caption{Error maps of ``boxer\_viewdep\_vox12''. Left: G-PCC (octree), bpp = 0.01, D1-PSNR = 56.80. Middle: G-PCC (trisoup), bpp = 0.02, D1-PSNR = 61.24. Right: HPSR-PCGC, bpp = 0.01, D1-PSNR = 62.57.}\label{fig:dense-error}
\end{figure}

\begin{figure}[!t]
    \centering
    \includegraphics[width=.75\linewidth]{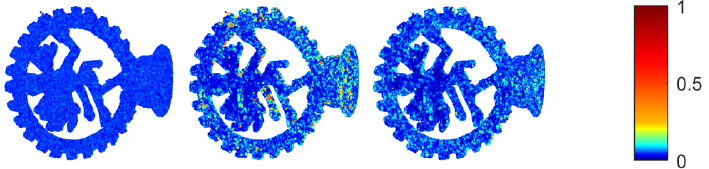}
    \caption{Error maps of ``shiva\_00035\_vox12''. Left: G-PCC (octree), bpp = 0.18, D1-PSNR = 58.92. Middle: G-PCC (trisoup), bpp = 0.29, D1-PSNR = 60.87. Right: HPSR-PCGC, bpp = 0.10, D1-PSNR = 60.85.}\label{fig:sparse-error}
\end{figure}

\noindent\textbf{Error map visualization}.
Figs.~\ref{fig:solid-error},~\ref{fig:dense-error}, and~\ref{fig:sparse-error} depict the error maps of three point clouds with different densities.  Careful visual inspection reveals distinct characteristics of different compression methods. For G-PCC (octree), the errors appear uniformly distributed, which may arise from grid downsampling. For G-PCC (trisoup), the presence of more sparsely distributed yellow and red regions in the error maps (such as the feet and fingers in ``basketball\_player\_vox11\_00000200'' as well as body parts in ``boxer\_viewdep\_vox12'') indicates more significant reconstruction errors in highly detailed regions. In contrast, HPSR-PCGC achieves better reconstruction quality using fewer bits than G-PCC (octree) and G-PCC (trisoup). 

\subsection{BDBR Comparison to SRLUT}
Table~\ref{tab:bdbr1} presents the D1- and D2-BDBR savings of HPSR-PCGC against SRLUT~\cite{borges2022fractional}. Note that we encountered memory limitations when generating SRLUT results for the point clouds ``facade\_00015\_vox14'' and ``ulb\_unicorn\_vox13'' at specific rate points, leading to missing data. HPSR-PCGC consistently outperforms SRLUT by a clear margin, as HPSR-PCGC encodes more accurate priors. 
The compared results further demonstrate the necessity of constructing the hierarchical prior at the encoder side for super resolution at the decoder side.

\begin{table}[!t]
\begin{center}
\caption{D1- and D2-BDBR savings of HPSR-PCGC against SRLUT} \label{tab:bdbr1}
\begin{tabular}{lcc}
  \toprule
  Point Cloud & D1 & D2 \\
  \midrule
  basketball\_player\_vox11\_00000200 & $-47.6\%$ & $-46.6\%$ \\
  dancer\_vox11\_00000001 & $-40.0\%$ & $-39.9\%$ \\
  facade\_00064\_vox11 & $-45.5\%$ & $-45.7\%$ \\
  longdress\_vox10\_1300 & $-29.6\%$ & $-31.4\%$ \\
  loot\_vox10\_1200 & $-33.7\%$ & $-30.6\%$ \\
  queen\_0200 & $-23.7\%$ & $-26.6\%$ \\
  redandblack\_vox10\_1550 & $-26.4\%$ & $-26.5\%$ \\ 
  soldier\_vox10\_0690& $-27.0\%$ & $-30.4\%$ \\ 
  thaidancer\_viewdep\_vox12 & $-46.9\%$ & $-48.9\%$ \\ 
  \midrule
  \textbf{Solid (Average)} & $-35.6\%$ & $-36.3\%$ \\ 
  \midrule
  boxer\_viewdep\_vox12 & $-62.1\%$ & $-64.2\%$ \\
  facade\_00009\_vox12 & $-44.1\%$ & $-48.3\%$ \\
  frog\_00067\_vox12& $-50.4\%$ & $-60.5\%$ \\
  head\_00039\_vox12& $-55.5\%$ & $-64.7\%$ \\
  house\_without\_roof\_00057\_vox12 & $-67.8\%$ & $-67.7\%$ \\ 
  longdress\_viewdep\_vox12 & $-53.2\%$ & $-56.5\%$ \\
  loot\_viewdep\_vox12 & $-55.8\%$ & $-60.3\%$ \\
  redandblack\_viewdep\_vox12 & $-44.4\%$ & $-50.5\%$\\ 
  soldier\_viewdep\_vox12 & $-50.9\%$ & $-56.9\%$ \\
  \midrule
  \textbf{Dense (Average)} & $-53.8\%$ & $-58.8\%$ \\
  \midrule
  egyptian\_mask\_vox12 & $-35.8\%$ & $-35.7\%$ \\
  shiva\_00035\_vox12 & $-28.3\%$ & $-33.9\%$ \\
  \midrule
  \textbf{Sparse (Average)} & $-32.0\%$ & $-34.8\%$ \\ 
  \midrule
  \midrule
  \textbf{All (Average)} & $-43.4\%$ & $-43.6\%$ \\ 
  \bottomrule
\end{tabular}
\end{center}
\end{table}

\begin{figure*}[!t]
\centering
\begin{minipage}[b]{0.3\linewidth}
  \centering
  \includegraphics[width=\linewidth]{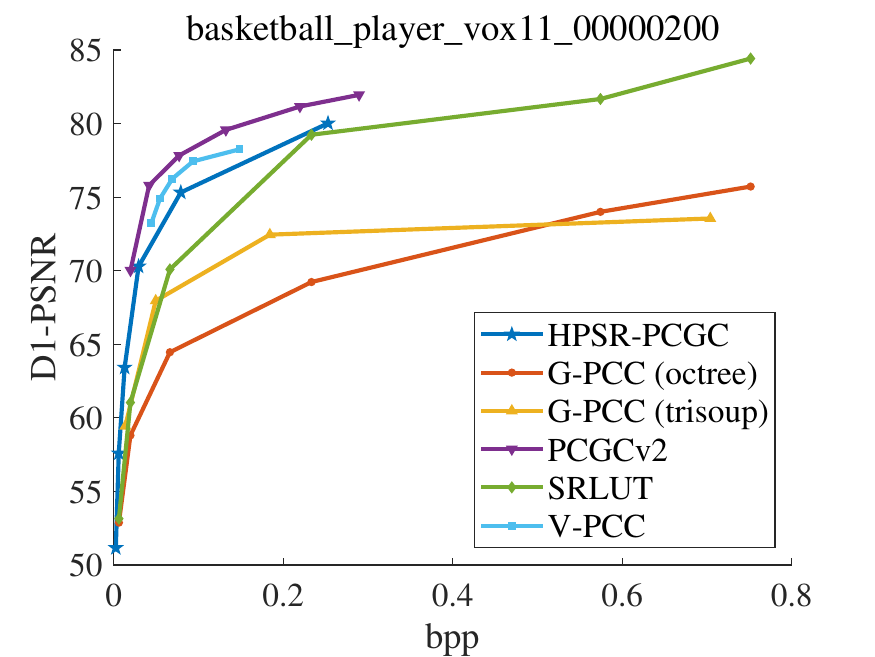}
\end{minipage}
\hspace{2mm}
\begin{minipage}[b]{0.3\linewidth}
  \centering
  \includegraphics[width=\linewidth]{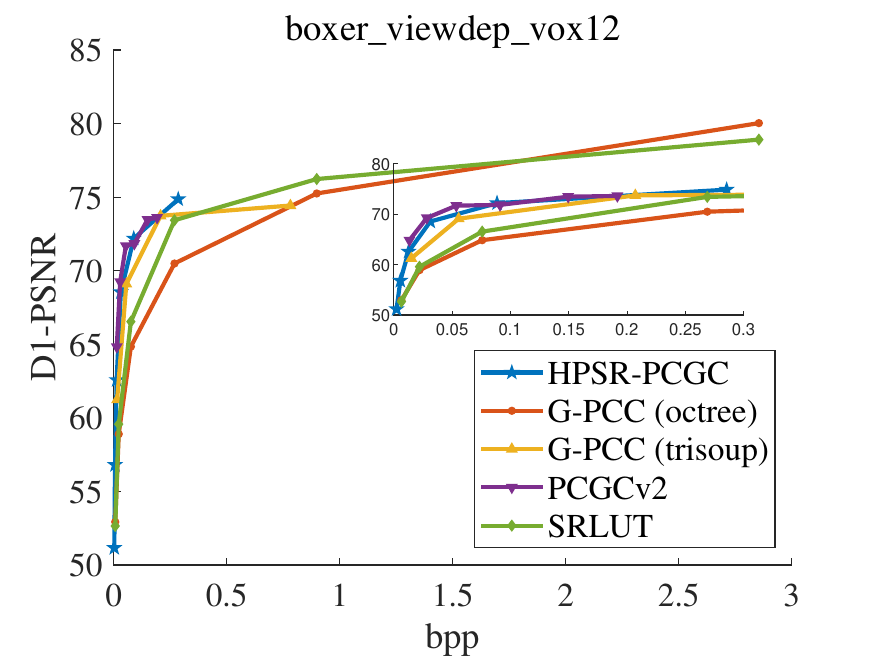}
\end{minipage}
\hspace{2mm}
\begin{minipage}[b]{0.3\linewidth}
  \centering
  \includegraphics[width=\linewidth]{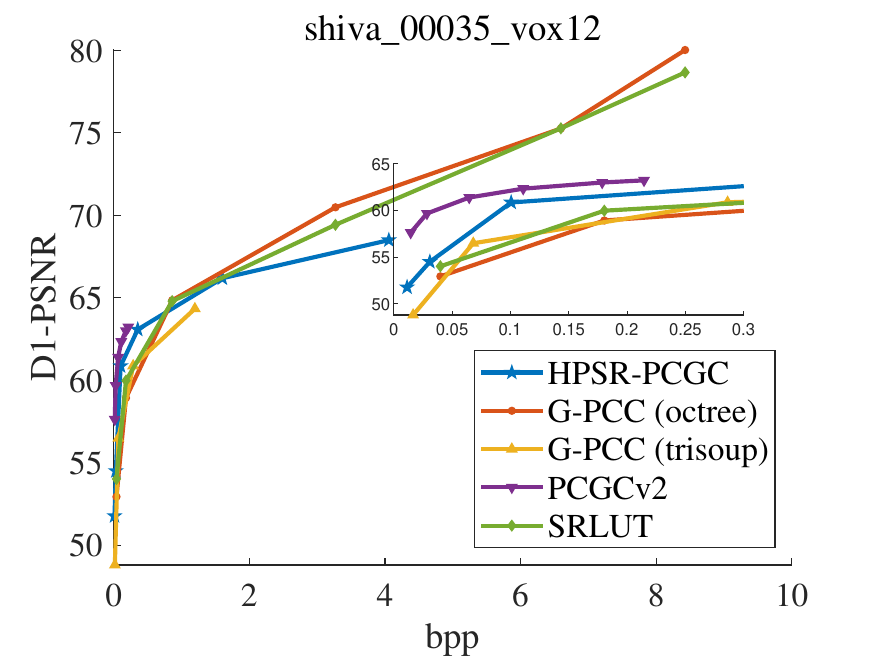}
\end{minipage}
\begin{minipage}[b]{0.3\linewidth}
  \centering
  \includegraphics[width=\linewidth]{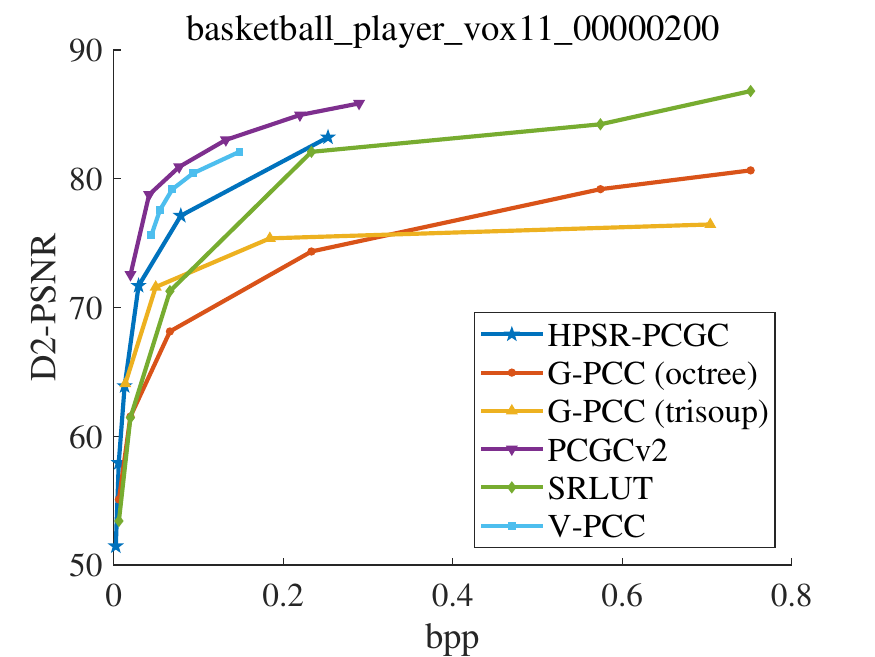}
\end{minipage}
\hspace{2mm}
\begin{minipage}[b]{0.3\linewidth}
  \centering
  \includegraphics[width=\linewidth]{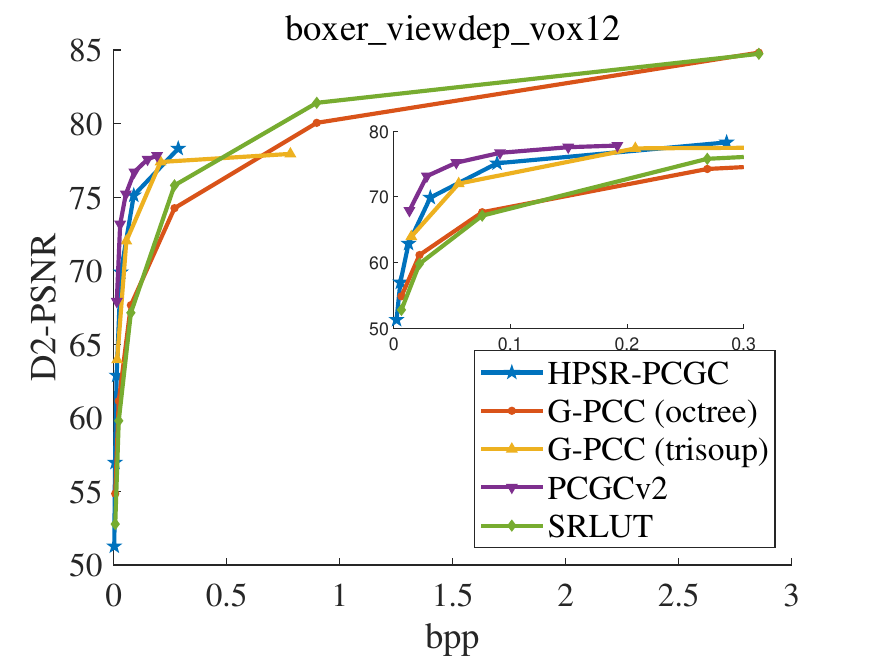}
\end{minipage}
\hspace{2mm}
\begin{minipage}[b]{0.3\linewidth}
  \centering
  \includegraphics[width=\linewidth]{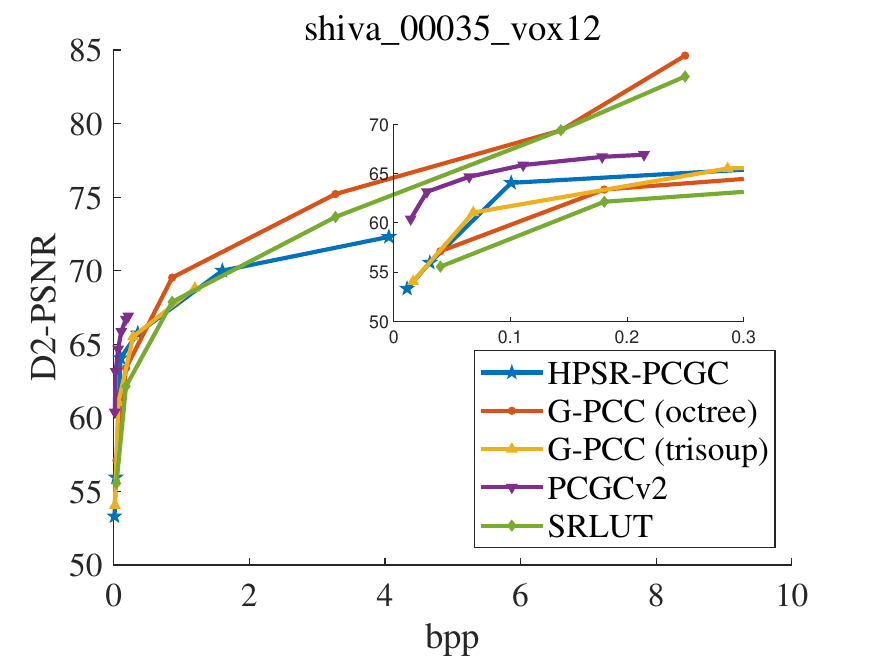}
\end{minipage}
\caption{Rate-distortion curves (\ie, bit per point [bpp] vs. D1-PSNR in the first row and D2-PSNR in the second row) for solid, dense, and sparse point clouds, respectively.}\label{fig:rd}
\end{figure*}

\subsection{BDBR Comparison to V-PCC and PCGCv2}
We also provide a reference comparison of HPSR-PCGC to V-PCC~\cite{VPCCv18} and the deep learning-based method PCGCv2~\cite{wang2021multiscale} in Table~\ref{tab:bdbr2}. 
V-PCC requires careful manual configuration of hyperparameters for each point cloud, which is time-consuming and challenging. We thus only tested V-PCC on seven solid point clouds using the suggested parameter configurations of their dynamic counterparts. HPSR-PCGC exhibits an average D1-BDBR overhead of $27.1\%$ to V-PCC. The performance gains of V-PCC are primarily due to the adoption of a mature video codec at the cost of longer encoding time, as shown in Subsec.~\ref{subsec:time}. PCGCv2, with no scaling, does not generalize well to dense and sparse point clouds, where HPSR-PCGC achieves an average of more than $90\%$ D1-BDBR savings. When we set the scaling factor to $1$, $0.375$, and $0.15$ for solid, dense, and sparse point clouds, respectively, this generalization issue of PCGCv2 is alleviated. Nevertheless, HPSR-PCGC still exhibits performance gains over PCGCv2 on sparse point clouds. 
Overall, HPSR-PCGC demonstrates substantial improvement over G-PCC v14, and closes the gap with V-PCC and PCGCv2 on solid point clouds, while inheriting the efficiency of G-PCC.

\begin{table}[!t]
\begin{center}
\caption{Average D1-BDBR of HPSR-PCGC against V-PCC and PCGCv2}\label{tab:bdbr2}
\begin{tabular}{lccc}
  \toprule
  Cat1A & V-PCC & PCGCv2 & PCGCv2 (no scaling) \\
  \midrule
  Solid & $27.1\%$ & $62.5\%$ & $62.5\%$ \\ 
  Dense & - & $28.9\%$ & $-96.0\%$\\ 
  Sparse & - & $-15.1\%$ & $-99.6\%$\\ 
  \bottomrule
\end{tabular}
\end{center}
\end{table}

\begin{table*}[!t]
    \centering
    \caption{Runtime comparison on point clouds with varying densities. ``$^*$'' indicates that the highest rate point has been reached and ``-'' indicates that the corresponding method is not applicable}
    \label{tab:runtime}
    \begin{tabular}{lccccccccc}
    \toprule
    \multirow{2}{*}{Method} & \multicolumn{3}{c}{Solid (basketball\_player\_vox11\_00000200)} & \multicolumn{3}{c}{Dense (boxer\_viewdep\_vox12)} & \multicolumn{3}{c}{Sparse (shiva\_00035\_vox12)} \\
    & bpp & D1-PSNR & Enc/Dec Time (s) & bpp & D1-PSNR & Enc/Dec Time (s) & bpp & D1-PSNR & Enc/Dec Time (s) \\
    \midrule
    G-PCC (octree) & 0.75 & 75.72$^*$ & 1.63/0.47 & 0.90 & 75.25 & 2.22/0.73 & 0.18 & 58.92 & 0.20/0.02 \\
    G-PCC (trisoup) & 0.70 & 73.55$^*$ & 5.11/3.25 & 0.21 & 73.73 & 4.00/2.45 & 0.29 & 60.87 & 1.95/0.39 \\
    SRLUT & 0.23 & 79.24 & 0.69/35.57 & 0.27 & 73.43 & 0.92/353.36 & 0.18 & 59.97 & 0.20/703.73 \\
    V-PCC & 0.15 & 78.25 & 104.2/3.83 & - & - & -/- & - & - & -/- \\
    PCGCv2 & 0.13 & 79.57 & 246.42/445.78 & 0.19 & 73.61 & 183.23/341.43 & 0.03 & 59.67 & 55.35/96.72 \\
    \midrule
    HPSR-PCGC & 0.25 & 80.02 & 3.02/0.55 & 0.29 & 74.86 & 9.00/1.50 & 0.10 & 60.85 & 1.31/0.48\\
    \bottomrule
    \end{tabular}
\end{table*}

\subsection{Rate-Distortion Curves}
The rate-distortion curves depicted in Fig.~\ref{fig:rd} provide valuable insights into different compression methods. For the solid point cloud ``basketball\_player\_vox11\_00000200'', HPSR-PCGC consistently outperforms G-PCC (octree), G-PCC (trisoup), and SRLUT across the entire PSNR range, confirming the effectiveness of our hierarchical prior. For the dense point cloud ``boxer\_viewdep\_vox12'', HPSR-PCGC achieves a significant improvement over G-PCC, approaching the performance of PCGCv2. Nevertheless, D1-/D2-BDBR values only reflect the bitrate savings within a specific D1-/D2-PSNR range, and the comparisons of G-PCC (octree) to other methods are only valid at lower bitrates. 
SRLUT fails to enhance G-PCC (octree) decoded point clouds at the highest rate point due to the violation of the cross-scale self-similarity assumption. For the sparse point cloud ``shiva\_00035\_vox12'', the rate-distortion curve of HPSR-PCGC is positioned below that of G-PCC (octree) at higher bitrates, but surpassing G-PCC (octree), G-PCC (trisoup), and SRLUT at lower bitrates. Although PCGCv2 achieves the best performance, it is only valid at a narrow range of very low bitrates. 

\subsection{Runtime Comparison}\label{subsec:time}
We compared the runtime of different methods using the same workstation equipped with an Intel Core i7-8700K CPU. Our implementation of G-PCC (octree), G-PCC (trisoup), V-PCC, and the proposed HPSR-PCGC, is written in C++. SRLUT is implemented using MATLAB, while PCGCv2 is implemented using PyTorch. We executed PCGCv2 in the CPU mode to ensure a fair comparison. Moreover, we tried to ensure the selected rate points of different methods are in a shared (narrow) D1-PSNR range, \eg, less than $2$ dB. The results are shown in Table~\ref{tab:runtime}. The encoding and decoding time of HPSR-PCGC is comparable to that of G-PCC, indicating that the added time complexity by hierarchical prior construction and hierarchical prior-based super resolution is marginal (relative to the achieved BDBR savings presented in Table~\ref{tab:bdbr}). As a post-processing method, the decoding time of SRLUT significantly increases. SRLUT can be accelerated by removing the data augmentation step at the cost of reduced performance, and the runtime should be faster if a C++ implementation is available. V-PCC is slow in encoding, which encompasses projection, patch packing, and video coding. PCGCv2 is even slower in encoding and decoding, due to the adoption of neural networks. Nevertheless, the runtime of PCGCv2 can be significantly reduced when the GPU mode is enabled, with a comparable encoding time and $10\times$ slower decoding time against G-PCC (octree) for solid point clouds.

\subsection{Bit Allocation Analysis}
Table~\ref{tab:bits} shows the bit allocation of HPSR-PCGC to the base point cloud $\V^{(K)}$ and the associated hierarchical prior $\{\bsigma^{(k)}\}_{k=1}^{K}$ on three point clouds, namely the solid ``basketball\_player\_vox11\_00000200'', the dense ``boxer\_viewdep\_vox12'', and the sparse ``shiva\_00035\_vox12''. As the bitrate increases, the bits used to encode $\V^{(K)}$ increase much faster than (and significantly surpass) the bits used to encode the prior. Up to r03, the hierarchical prior consumes more bits than the base point cloud. This arises because the base point cloud is solid in these bitrates and compactly compressed with G-PCC (octree), while the coding of the hierarchical prior is not optimized in HPSR-PCGC.

\begin{table}[!t]
\begin{center}
\caption{Bit allocation analysis of HPSR-PCGC} \label{tab:bits}
\resizebox{\linewidth}{!}{
\begin{tabular}{lcccccc}
  \toprule
  \multirow{2}{*}{Rate} & \multicolumn{2}{c}{Solid} & \multicolumn{2}{c}{Dense} & \multicolumn{2}{c}{Sparse} \\
  & $\V^{(K)}$ & $\{\bsigma^{(k)}\}$ & $\V^{(K)}$ & $\{\bsigma^{(k)}\}$ & $\V^{(K)}$ & $\{\bsigma^{(k)}\}$ \\
  \midrule
  r01 & 1,712 & 5,408 & 2,096 & 6,328 & 3,112 & 8,424 \\
  r02 & 5,560 & 11,704 & 6,960 & 12,752 & 10,208 & 21,016 \\
  r03 & 17,704 & 19,864 & 22,952 & 21,400 & 40,360 & 61,224 \\
  r04 & 57,648 & 28,120 & 76,872 & 32,632 & 181,952 & 174,920 \\
  r05 & 193,296 & 38,144 & 264,744 & 44,560 & 871,280 & 745,136 \\
  r06 & 682,408 & 57,104 & 940,248 & 56,464 & 3,298,328 & 795,360 \\
  \bottomrule
\end{tabular}
}
\end{center}
\end{table}

\begin{figure*}[!t]
\centering
\begin{minipage}[b]{0.27\linewidth}
  \centering
  \includegraphics[width=\linewidth]{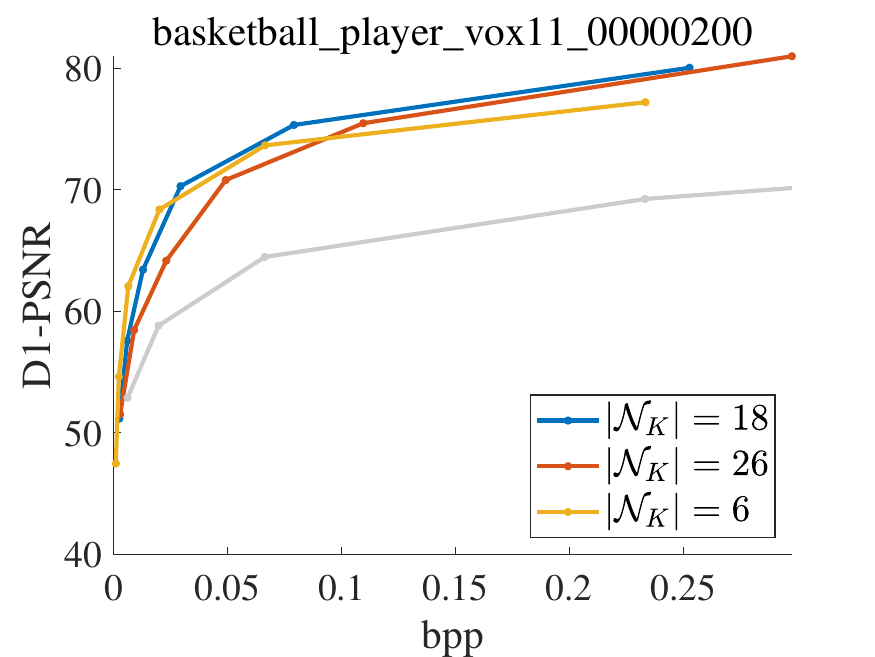}
\end{minipage}
\hspace{2mm}
\begin{minipage}[b]{0.27\linewidth}
  \centering
  \includegraphics[width=\linewidth]{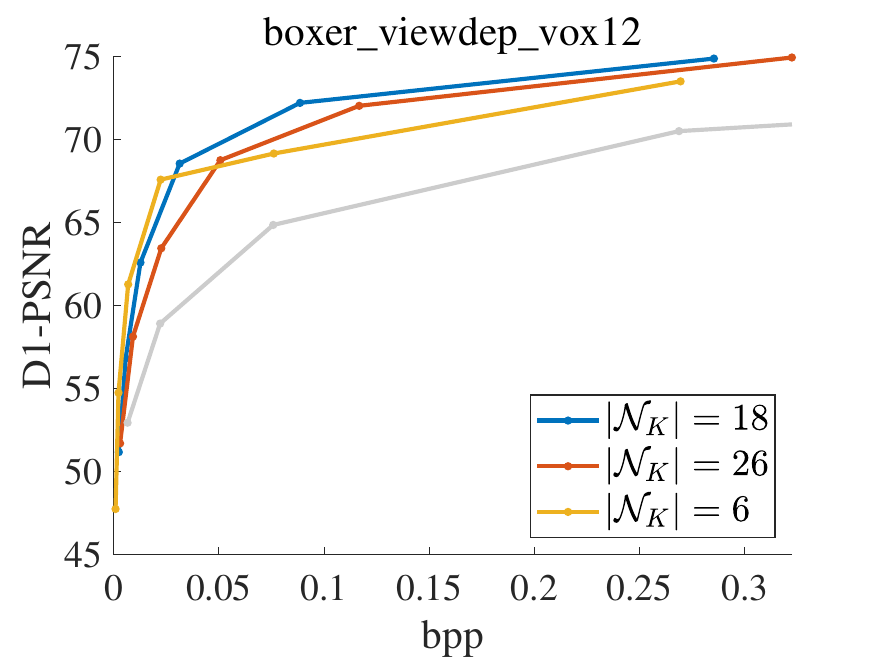}
\end{minipage}
\hspace{2mm}
\begin{minipage}[b]{0.27\linewidth}
  \centering
  \includegraphics[width=\linewidth]{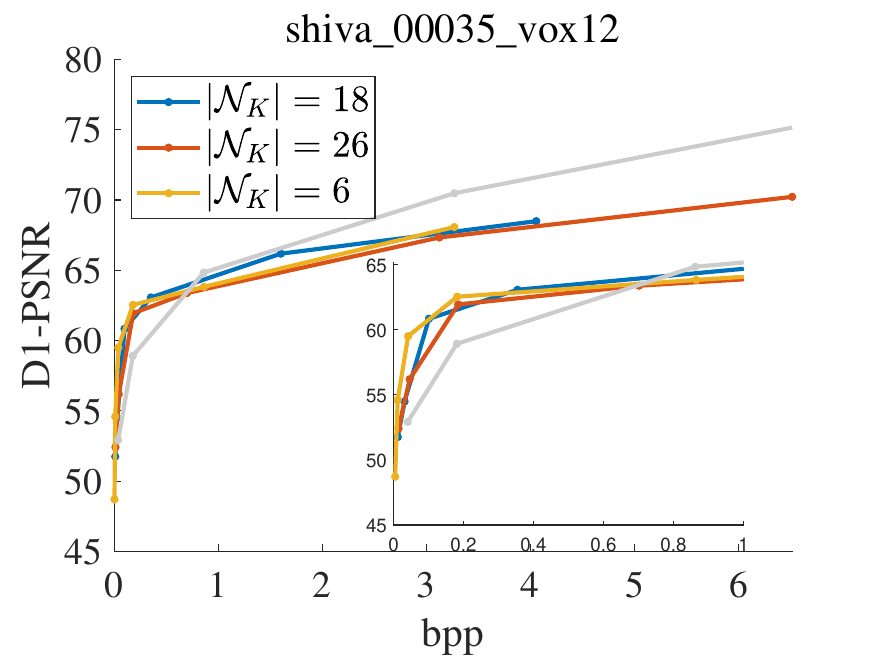}
\end{minipage}
\caption{Rate-distortion curves (\ie, bpp vs. D1-PSNR) of HPSR-PCGC by varying the number of neighbors. The gray curve represents G-PCC (octree). Local details are enlarged in ``shiva\_00035\_vox12''.}\label{fig:rd-prior}
\end{figure*}

\subsection{Discussion}
\noindent\textbf{Choices of $K$ and $K'$}.
Although the proposed HPSR-PCGC offers improved time complexity compared to V-PCC and deep learning-based approaches, it still lags behind octree-based G-PCC, particularly in terms of the encoding time. Nevertheless, the encoding time complexity can be reduced by adjusting the hyperparameters such as decreasing the number of neighbors and the value of $K$ during hierarchical prior construction. The decoding time complexity can be further optimized as well by reducing the number of interpolations, $K+K'$. For instance, by setting $K'=0$ (\ie, skipping the super resolution of $\hV^{(0)}$), the decoding time can be reduced to $83\%$ of G-PCC (octree) while still achieving an average of $61.7\%$ D1-BDBR savings and $43.0\%$ D2-BDBR savings on the MPEG Cat1A dataset. 

\noindent\textbf{Choices of $\mathcal{N}_K$ and prior coder}.
The accuracy of the hierarchical prior is directly influenced by the number of neighbors considered in $\mathcal{N}_{K}$. Fig.~\ref{fig:rd-prior} depicts the rate-distortion curves of HPSR-PCGC with different numbers of neighbors. $\vert\mathcal{N}_{K}\vert$ equals $6$, $18$, and $26$ corresponding to voxel neighbors with shared faces, lines, and vertexes, respectively.  More neighbors generally lead to better reconstruction quality for the same rate point. As there is no free lunch in data compression, the more accurate prior requires more bits for encoding. We find that $18$ neighbors yield the best trade-off.

A more promising way of determining $\mathcal{N}_K$ is through Rate-Distortion Optimization (RDO). Here, we conduct preliminary exploration, where we adaptively determine the local neighbors based on the cost for encoding the base point cloud $\V^{(K)}$. Fewer local neighbors are considered if the cost for encoding $\V^{(K)}$ is smaller. Besides, we adopt the standard arithmetic coding to further compress the hierarchical prior. We denote this implementation as HPSR-PCGC-RDO, and more details can be found at \url{https://github.com/lidq92/mpeg-pcc-tmc13/tree/hpsr_pcgc_rdo}. From the right side of Table~\ref{tab:bdbr}, we find that HPSR-PCGC-RDO offers approximately $15\%$ more BDBR savings than HPSR-PCGC compared to G-PCC (trisoup) on the MPEG Cat1A dataset. This verifies the effectiveness of the adaptive selection of local neighbors and the arithmetic coding in the prior coder. 

\section{Conclusion and Future Work}\label{sec:conclusion}
We have introduced a hierarchical prior for lossy point cloud geometry compression. The hierarchical prior is constructed during encoding, which serves as side information for coarse-to-fine super resolution of the point cloud during decoding. Our experimental results demonstrate significant D1-/D2-BDBR savings while maintaining acceptable time complexity across point clouds with varying densities compared to G-PCC. Our current work focuses solely on lossy geometry coding, while several potential directions are worth exploring.

\noindent\textbf{Further BDBR savings}.
The proposed HPSR-PCGC  underperforms V-PCC and deep learning-based PCC for solid point clouds. Currently, the hierarchical prior construction relies on simple frequency-based statistics, which could be replaced by learnable computational modules like neural networks to achieve improved rate-distortion performance. Additionally, density-adaptive techniques could be integrated into HPSR-PCGC to better accommodate point clouds with different densities. For instance, we could employ a lightweight neural network to estimate the point cloud density, and set appropriate hyperparameters adaptively. These techniques together may encourage beneficial early stopping when interpolating sparse point clouds. 

\noindent\textbf{Joint compression of geometry and attributes}.
Since point clouds are often associated with attributes such as color, reflectance, and surface normal, it is crucial to jointly compress point cloud geometry and attributes. A na\"{i}ve extension of HPSR-PCGC to recoloring newly interpolated points is to inherit the attributes from their nearest colored points. However, this method may be ineffective in reconstructing attributes of significant variations. Similar to geometry coding, (hierarchical) priors for attribute enhancement can be constructed using computational methods such as Wiener filtering~\cite{wiener1949extrapolation} and other learnable modules~\cite{sheng2022attribute,wang2022cu}.

\noindent\textbf{Near-lossless and lossless compression}.
The proposed hierarchical prior has the potential to be extended to near-lossless and lossless point cloud geometry compression. One possible implementation is to also encode the residuals, which capture the discrepancies between the interpolated point cloud and the original point cloud~\cite{li2022near}.


\bibliographystyle{IEEEtran}
\bibliography{IEEEabrv,PCC}

\end{document}